\documentclass[useAMS,usenatbib]{mn2e}

\usepackage{graphicx}
\usepackage{epstopdf}
\usepackage{amsfonts}
\usepackage{amssymb}
\usepackage{amsmath}

\usepackage{color}

\title[Evolution of Ly$\alpha$ Blobs]
  {Modelling the Evolution of Ly$\alpha$ Blobs and Ly$\alpha$ Emitters}
\author[M. Smailagi{\' c} et al.]
  {M.~Smailagi{\' c},$^1$\thanks{E-mail:marijana@aob.rs}
  M.~Micic,$^1$ and N.~Martinovi{\' c}$^1$\\
  $^1$Astronomical Observatory Belgrade, Volgina 7, 11060 Belgrade, Serbia\\}
\date{18 February 2016}   %%{Released 2014 Xxxxx XX}

\pagerange{\pageref{firstpage}--\pageref{lastpage}} \pubyear{2015}

\def\LaTeX{L\kern-.36em\raise.3ex\hbox{a}\kern-.15em
    T\kern-.1667em\lower.7ex\hbox{E}\kern-.125emX}

\begin{document}

\label{firstpage}

\maketitle

\begin{abstract}
In this work we model the observed evolution in comoving number
density of Lyman-alpha blobs (LABs) as a function of redshift, and try to
find which mechanism of emission is dominant in LAB.
Our model calculates LAB emission both from cooling radiation from the
intergalactic gas accreting onto galaxies and from star
formation (SF). We have used dark matter (DM) cosmological
simulation to which we applied empirical recipes for Ly$\alpha$
emission produced by cooling radiation and SF in every halo. In difference to the
previous work, the simulated volume in the DM simulation
is large enough to produce an average LABs number density.
 At a range of redshifts $z\sim 1-7$ we compare our results with the observed luminosity functions of
 LABs and LAEs.
Our cooling radiation luminosities appeared to be too small to explain
LAB luminosities at all redshifts. In contrast, for SF we obtained
a good agreement with observed LFs at all redshifts studied.
We also discuss uncertainties which could influence the obtained results,
and how LAB LFs could be related to each other in fields with different density.
\end{abstract}

\begin{keywords}
 %% \LaTeXe\ -- class files: \verb"mn2e.cls"\ -- sample text -- user guide.
%% galaxies: evolution -- galaxies: high-redshift -- galaxies: luminosity function, mass function -- galaxies: star formation
   galaxies: evolution -- galaxies: high-redshift -- galaxies: luminosity function, mass function -- galaxies: stellar content
   -- (galaxies:) cooling flows

\end{keywords}

\section{Introduction}\label{sec:1}

 Ly$\alpha$ blobs (LABs) are very luminous ($\sim
10^{43}$-$10^{44}$ erg s$^{-1}$) and very extended (with diameters
of $\sim 50$-$100$ kpc and more) regions of Ly$\alpha$ emission,
which are radio quiet. They are observed at a range of redshifts
$z \sim 1-6.6$, but the bulk of objects currently known is found
between $z \sim 2-3$ \citep[e.g.][]{Steidel00,M04,M11,Sai06,Ou09,Y09,Y10,E11,Bridge13,Pr12}.
Similar objects are large
Ly$\alpha$ nebulae surrounding some high redshift radio galaxies \citep[e.g.][]{McCa87,VillM07}.
However, it is much less clear what is the source of emission in LABs.
Most probable proposed sources of energy are: photoionization by starbursts or by active galactic
nuclei (AGN) \citep[e.g.][]{Gea09}, superwinds driven by starburst
supernovae \citep[e.g.][]{TaShi00}, cooling radiation from cold
streams of gas accreting onto galaxies \citep[e.g.][]{Hai00,DL09,G10}
, or some combination thereof
\citep[e.g.][]{Col11}. LABs are rare (with comoving number density
$\sim 10^{-6}$ Mpc$^{-3}$ - $10^{-4}$ Mpc$^{-3}$) and
preferentially found in overdense regions \citep[e.g.][]{Y10},
which indicates that LABs could be the sites of the formation of
the most massive galaxies. LABs comoving number density decreases
at low redshifts \citep[see][]{Keel09,Pr09,Barg12}.
LABs are not observed below $z\lesssim 1$, but,
in the local Universe some star-forming galaxies are observed with
extended Ly$\alpha$ emission to $\sim 10-30$ kpc \citep{Keel05,Hay13}.

 A large number of LABs is associated with submillimetre and
infrared sources \citep{Chap01,Gea05,Gea07} or with
obscured AGN \citep{BaZy04,ScaCol09}, which could indicate a
significant role of stellar feedback and AGN in their luminosity.
For example, \citet{Col11} used mid-IR and
submillimeter observations of LABs at $z=2.38-3.09$ which are
detected by \citet{M04} and \citet{Pa04}, and they have found that IR
emission of 60 per cent of LABs originates mostly from SF, while
the rest are powered by AGN or extreme starburst.
In addition, a large LAB firstly discovered by \citet{Steidel00}
is observed to have polarized radiation, which indicates presence
of central source of energy \citep{Hayes11}.
 However, a number of LABs are not associated with sources that are
powerful enough to explain the observed Ly$\alpha$ luminosities, or
have large equivalent widths which are not
easily explained with star formation. In such
cases cooling radiation could play a dominant role in LAB luminosity
\citep{N06,M06,Sm08,Sai08} \citep[but see also][]{Pr15}.
For example, \citet{M04} found that $1/3$ of the LABs which they
observed at $z=3.1$ have too large equivalent widths (EWs) to be
explained with simple photoionization by massive stars with %% in which
Salpeter initial mass function. However, these LABs   %%is assumed
could still be explained if the stars are zero-metallicity stars
or with stellar initial mass function with an extreme slope of
$\alpha = 0.5$, or if ionizing UV sources are hidden from our line
of sight. In addition, $\sim 40$ percent of objects observed by
\citet{Sai08} at $z\sim3-5$ are most likely powered by cooling radiation,
%%they have very large EWs which cannot be explained with photoionization
%%by a moderately old ($>10^{7}$ yr) stellar population, they show
%%no clear signature of AGN and superwinds,
and they show a correlation between Ly$\alpha$ luminosities and velocity widths.
Besides these sources of energy, resonant scattering could significantly influence the observed
Ly$\alpha$ emission. \citet{Steidel11} observed SB profiles in
Ly$\alpha$ line and UV continuum in deep imaging. They found that
the Ly$\alpha$ emission comes mainly from scattered radiation from
galaxy HII regions, and that on average the contribution from cooling radiation
is not significant.

Similar, but much more common and usually less bright, objects are Lyman-alpha emitters (LAEs).
They are usually defined as objects with Ly$\alpha$ equivalent width larger than
20\AA. \citet{Steidel11} observed LAEs at $z\sim 2-3$, and
concluded that if observations are deep enough than all LAEs would
be classified as LABs, with extended Ly$\alpha$ emission \citep[but see also][]{Hay13,Wis15}.
 LAEs are observed at a range of redshifts from $z\sim 0$ to $z\sim 7$, and higher
 \citep[e.g.][]{Dawson07,Ou08,Barg12}. Most of
LAEs at $z\sim 2-7$ are young metal-poor star-forming galaxies,
most probably with a negligible fraction of AGN activity \citep[e.g.][]{Gaw06e,Fin07e}.
Some LAEs have high equivalent
widths, which are not explained by a simple star formation with a Salpeter IMF.
As determined from population synthesis models, LAEs
have small stellar masses of $10^{8}-10^{9} M_{\odot}$
\citep[e.g.][]{Nil07e}. At lower redshifts, $z\sim 1$, AGN
activity is detected in some fraction of LAEs \citep{Barg12}.

\vspace{12pt}

Observations and theoretical models showed that the accretion of
gas from the intergalactic medium has an important role in galaxy
formation and evolution. Numerical simulations show that galaxies
acquire their gas through the cold and hot modes \citep[e.g.][]{Katz03,Keres05,Keres09,Ocv08,Dek09}.
According to the simulations, in the hot mode, the
accreting gas is shock heated to roughly the virial temperature.
After cooling it collapses into galaxies in a presumably
spherically symmetric manner.
 In the cold mode, gas maintains a temperature of $T<2.5 \times 10^{5}$ K and is
accreted onto galaxies in the form of filamentary streams.
Simulations show that most of the baryons in a galaxy are accreted
via the cold mode \citep[e.g.][]{Keres09}. While the cold gas is
streaming towards the dark-matter halo potential well,
gravitational binding energy is released and the hydrogen atoms
are excited, followed by cooling emission of Ly$\alpha$ \citep[e.g.][]{Hai00,Furl05,DekBir08}.

Previously a number of authors have created simulations and
analytical models which tried to explain LAB emission through the
cooling radiation alone. Some of them are briefly summarized here.
\citet{DL09} developed an analytical model to predict Ly$\alpha$
emission of galaxies from cooling radiation of cold gas accreting into galaxies.
According to their work, if $\gtrsim 20$ per cent of the
gravitational energy of the gas is radiated away, their LABs have
similar Ly$\alpha$ luminosity functions and line widths as the
observed LABs at $z=3.1$, however their diameters seem to be too
large. \citet{G10} used cosmological hydrodynamical simulations to
calculate Ly$\alpha$ emission from cooling radiation in galaxies inside massive
haloes at high redshift, where LABs are expected to be found.
Their luminosity function is in agreement with observation of
\citet{M04} at $z=3.1$, and the relation luminosity -- area is
roughly in agreement with the same observation. In addition, they
derived an analytical model based on released gravitational energy
from infalling gas, which provides similar results. However,
\citet{FG10} showed that adding a more precise calculation of
radiative transfer changes these results. They used cosmological
hydrodynamical simulations to predict Ly$\alpha$ emission from cooling radiation,
which included a precise calculation of radiative transfer of the
Ly$\alpha$ emission. When self-shielding and sub-resolution models
are properly included in a simulation, the computed luminosities
could differ by an order of magnitude. \citet{FG10} concluded that
it is difficult to explain LABs luminosities only with cooling radiation, unless
if the gas of density sufficient to form stars is in gaseous
phase. \citet{Ros12} simulated massive haloes and described cold
streams with better resolution than in previous simulations. They
have also properly modelled self shielding from UV background, and
calculated gravitational efficiency of cooling radiation. Their simulated LABs
have luminosity, extent and morphology in agreement with
observations, but they slightly overpredict LABs abundances (at
$z=3$).

Photoionization by starbursts alone or in a combination with cooling radiation is
another probable model proposed to explain LABs \citep[e.g.][]{Furl05,Cen12}.
Recently, \citet{Cen12}
developed a starburst model for LABs, in which they also included
Ly$\alpha$ emission from cooling radiation. In their model, emission from
gravitational sources (which includes cooling radiation) is significant, but
sub-dominant compared to stellar emission. They successfully
reproduced LAB luminosity function and luminosity-size relation at
$z=3.1$. \citet{Yaj12} used cosmological hydrodynamic simulations
to predict Ly$\alpha$ properties of progenitors of local $L^{*}$
galaxies with size and substructure similar to Milky Way.
According to their results, Ly$\alpha$ emission from cooling radiation increases
with redshift, contributing roughly $50$ per cent of the total at
$z \gtrsim 6$.

\vspace{12pt}

However, in all of these works derived properties of simulated
LABs were compared to the observed LABs at $z=3.1$ (\citet{M11}
survey, volume $1.6 \times 10^{6} {\rm Mpc^{3}}$), but none of
them included observations at other redshifts. More importantly,
their volumes are less than a volume necessary to produce an
average LABs number density: e.g. \citet{M11} survey has a volume
of $1.6 \times 10^{6}$ Mpc$^{3}$, while simulations have volumes
of up to $\sim 10^{5}$ Mpc$^{3}$ (Table \ref{tab:1}).
An exception is analytical model of \citet{DL09} who assumed Sheth-Tormen distribution of haloes and \citet{NFW97} dark matter profiles.
It is computationally expensive to simulate a large volume in a
hydrodynamical simulation and keep all of the required physics. In
an another approach, large scale cosmological dark matter (DM)
simulations could be combined with semi-analytical recipes. This
is applied in our work.
 The recipes used include cold gas accretion rates from a
(smaller-scale) hydrodynamical simulation, stellar masses from
matching of DM haloes to the observed galaxies, escape fraction of
Ly$\alpha$ photons from comparison of observed SFR functions to
Ly$\alpha$ luminosity functions, and intergalactic opacity from
observations of Ly$\alpha$ forest.

\begin{table*}
\begin{minipage}{150mm}
\caption{Comoving volumes and maximal halo virial masses (at
$z\sim 3$) in selected simulations.
} \label{symbols}
\begin{tabular}{@{}lcc}
\hline Reference            & $V_{\rm com} [10^{6} Mpc^{3}]$ & $M_{\max} [M_{\odot}]$  \\
\hline \citet{G10} & 0.02, 0.36 & $10^{13}$ \\
     \citet{FG10}  & 0.19 & $2.5 \times 10^{11}$ \\
     \citet{Yaj12} & 0.00036 & $1.6 \times 10^{12}$ \\
     \citet{Ros12} & 0.13 & $1.3 \times 10^{13}$ \\
     \citet{Cen12} & 0.03 & $5 \times 10^{12}$ \\
\hline present work & 6.4 & $2.5\times 10^{13}$ \\
\hline
\end{tabular}
\medskip
\label{tab:1}
\end{minipage}
\end{table*}

With this, for every DM halo we calculate Ly$\alpha$ luminosities
from cooling radiation and from SF. At a range of redshifts $z\sim1-7$ we
determine luminosity functions, and compare them with the
observations in order to determine which mechanism is the dominant
source of Ly$\alpha$ emission in LABs. The outline of this paper
is as follows. In section {\S}\ref{sec:2} we describe the dark matter simulation.
In {\S}\ref{sec:3} we summarize observations at different redshifts.
Section {\S}\ref{sec:4} describes the physical models used, which
include computation of Ly$\alpha$ luminosity from cooling radiation and SF.
Our results are presented in section {\S}\ref{sec:5}. In
{\S}\ref{sec:5.1} it is shown how our calculated LAB luminosities depend
on halo masses.
Our luminosity functions are compared with observations
in sections {\S}\ref{sec:5.2} and {\S}\ref{sec:5.3}.
The calculated star formation rate functions
are compared with observations in {\S}\ref{sec:5.5}.
A discussion about our model is presented in section {\S}\ref{sec:6}, and includes
LAB areas and influence of different parameters
on the results. Our conclusions are summarized in section {\S}\ref{sec:7}.
Details about cold gas accretion rates and influence of the resolution of the DM simulation
are presented in Appendix {\S}\ref{sec:X}.

\section{Dark matter simulation}\label{sec:2}

Cosmological simulation used in this paper evolves a $\Lambda CDM$
cosmology in a periodic box with a side of $130 \,\,
\rm{h^{-1}Mpc}$ and a $512^{3}$ dark matter particles which
translates to a particle mass resolution of $1.14 \times
10^{9}\,\,\rm{h^{-1} M_{\odot}}$.

Initial conditions were configured using LasDamas cosmology
\citep{McBridge} which assumes flat geometry and cosmological
parameters with values: $\Omega_{m} = 0.25$, $\Omega_{\Lambda} =
0.75$, $\Omega_{b} = 0.04$, $h = 0.7$. Linear expansion part of
the simulation was computed using 2nd order Lagrangian
perturbation theory \citep{Crocce,Scoccimarro}.
Transfer function was calculated by the CMBfast code
\citep{Seljak} with assumed power-law index of $n_{s} = 1$ for
primordial power spectrum and rms mass fluctuations on $8 \,\,
\rm{h^{-1}Mpc}$ scale of $\sigma_{8} = 0.8$.

Simulation was executed with GADGET2 code \citep{Spr05} starting
from $z = 599$ and with a softening length of $\epsilon = 8 \,\,
\rm{h^{-1}kpc}$. Groups were first found by friends-of-friends
(FOF) code with a linking length of $b = 0.2$ of a mean particle
separation. After that, SUBFIND algorithm \citep{Spr01} was used
to find subhaloes within the groups by defining them as self-bound
locally overdense structures. For identification of dark matter
subhaloes we use resolution of at least 50 bound particles, which
corresponds to haloes with mass of $\sim 10^{11} \,\,
\rm{h^{-1}M_{\odot}}$.

\section{Observations}\label{sec:3}

Our results are compared with LABs and LAEs observations at $z=1-6.6$.
In order to do this, we found all surveys
in which $\geq 1$ LABs are observed ($> 1$ LABs at $z\sim 2-3$).
Properties of these surveys are summarized in Table \ref{tab:2}.
We present survey redshifts,
comoving volumes, number density contrasts (usually in the number of LAEs), %%or in DM,
surface brightness (SB) thresholds which would correspond to
$z=3.1$ (denoted with $SB(z=3.1)$), and number of LABs detected.
The corresponding $SB(z=3.1)$
are calculated from $SB(z=3.1)/SB(z)=(z+1)^{4}/(3.1+1)^{4}$.
In Table \ref{tab:3} we summarize observations of LAEs with which we compare our results.

We determine observed cumulative
luminosity functions directly from published LABs
luminosities in different surveys, with exception of
\citet{Sai06} and \citet{M11}.
During this, for \citet{Y09} survey we use their
LAB number densities which are corrected for incompleteness.
For \citet{M11} %%observations %%($z=3.1$)
we use their cumulative luminosity function from \citet{G10}
and cumulative area function from \citet{Ros12}.
For \citet{Sai06} we determine cumulative luminosity function
from their non-cumulative luminosity function \citep[fig. 13 in][]{Sai06},
%% in their fig. 13,
and compare with our results at $z=4-5$.

\begin{table*}
\begin{minipage}{150mm}
\caption{Properties of LABs surveys: survey reference, redshift
($z$), comoving volume ($V_{\rm com}$), surface brightness threshold which would
correspond to $z=3.1$ (${\rm SB}_{\rm th}$), number of LABs
detected ($n_{\rm LABs}$), and , number density contrast of
LAEs ($\delta$).} \label{symbols}
\begin{tabular}{@{}lcccccc}
\hline Survey           & $z$ & $V_{\rm com} $ & ${\rm SB}_{\rm th}$ & $n_{\rm LABs}$ & $\delta$ &  \\
  &  & $[10^{6}Mpc^{-3}]$ & ${\rm [erg s^{-1} cm^{-2} arcsec^{-2}]}$ & & &  \\
    \hline \citet{Keel09} & 0.8 &   $\sim2$  &  0.3    &  0  &   $>0$ $^{*3}$ &   \\
           \citet{Barg12} & 1   &   2.3      &  0.4    &  1  &   0 &  \\
           \citet{Pr09}   &1.6-2.9& 130      &7 $^{*2}$ &  5  &   0 &  \\
              \citet{Y09} & 2.3 &   2.14     & 2.1     &  4  &   0 &  \\
\citet{Y10} (fields) $^{*1}$  &2.3 &   0.44     & 2.3     &  9 &   0 &       \\
       \citet{Y10} (CDFS) & 2.3 &   0.11     & 2.3     &  16 &   5.7 $^{*4}$ &    \\
              \citet{E11} & 2.3 &   0.1      & 0.6     &  6  &    7 $^{*5}$ &  \citep{Steidel05}   \\
              \citet{M04} & 3.1 &   0.14     & 2.2     &  35 &   4.7 &   \citep{Steidel00}   \\
              \citet{M11} & 3.1 &   1.6      & 1.4     & 201 &   $\sim 0$ $^{*6}$  &     \\
            \citet{Sai06} &3.24-4.95& $\sim3$ &$\sim 8$&  41 &   0 &    \\
             \citet{Ou09} & 6.6 &   0.82     &  11.8   &  1  &   1 & \citep{Ou09}  \\
%%   Prescott et al. 2008. & 2.7 &   0.43    &   2 $\delta_{LAE}=2$, intermediate-band filter    1  nn   \\
%%    Matsuda et al. 2011. & 3.1 &   0.74    &   3.36? $\delta_{LAE}=5, 3.36$            nn?     \\
\hline
\end{tabular}
\medskip
\label{tab:2}

$^{*1}$ - fields CDF-N, COSMOS1, COSMOS2

$^{*2}$ -  In difference to most of the other work, \citet{Pr09} used broad-band survey to observe LABs. As \citet{Pr09} mentioned, their SB threshold is similar to \citet{Y09}. %% at $z\sim1.6-2.9$.

$^{*3}$ - these authors observed clusters of galaxies, however
they did not discuss their number density contrast

$^{*4}$ - We have estimated $\delta \sim 5.7$ of the peak in
the LAE number density from the figure 12 in \citet{Y10}, with the assumption
that the minimum surface density contour corresponds to the
average LAE number density.

$^{*5}$ - $\delta \sim 7$ is the number density contrast of galaxies.
\citet{Steidel05} have also determined density contrast in dark matter, $\delta_{DM}\sim1.8$. %%  \approx1.8$.}

$^{*6}$ - see \citet{G10} and \citet{Ros12}

\end{minipage}
\end{table*}

\begin{table*}
\begin{minipage}{150mm}
\caption{Properties of LAEs surveys: survey reference and redshift
($z$).} \label{symbols}
\begin{tabular}{@{}lcccc}
\hline Survey            & $z$ \\
\hline \citet{Barg12}    & 0.67-1.16 \\
     \citet{Hayes10}     & 2.2 \\
     \citet{Kudritzki00} & 3.1 \\
     \citet{Dawson07}    & 4.5 \\
     \citet{Taniguchi05} & 6.6 \\
\hline
\end{tabular}
\medskip
\label{tab:3}
\end{minipage}
\end{table*}

\section{Method}\label{sec:4}

In our work a large scale cosmological dark matter (DM)
simulations is combined with semi-analytical recipes.
The simulation is described in \citet{NM14} and in our section {\S}\ref{sec:2}.
In this section we describe
 how we compute emission from cooling
radiation ({\S}\ref{sec:4.1}) and from star formation
({\S}\ref{sec:4.2}).
Section {\S}\ref{sec:4.3} explains how we determine mean luminosity functions.

\subsection{Cold mode gas accretion}\label{sec:4.1}

%%Streams of cold gas have temperatures $T \sim 10^{4}-10^{5}{\rm
%%K}$. At these temperatures radiation is primarily emitted in the Ly$\alpha$ line.
%%While cold gas is accreting from $R_{\rm vir}$ toward the centre
%%of a halo, gravitational potential energy is released, and H atoms
%%are excited or ionized due to collisions. Ly$\alpha$ is emitted
%%during de-excitation or recombination. The Ly$\alpha$ emissivity
%%from these processes depends on temperature and on density and ionization state of gas.
In this model Ly$\alpha$ emission is calculated from the released
gravitational potential energy.
While cold gas is streaming from virial radius $R_{\rm vir}$ to
some radius $r_{0}$ in a halo, released gravitational potential
energy per unit time between radii $r_{1}$ and $r_{2}$ is equal to

  \begin{equation}
    \label{eq:7}
 {%
    {\dot{E}_{\rm grav}}(r)=\dot{M_{\rm c}}(r)\left|\frac{\partial\Phi}
    {\partial r}\right| ,
    }
%  \overfullrule 5pt
%  \mathindent\linewidth\relax
%  \advance\mathindent-259pt
  \end{equation}
where $r=(r_{1}+r_{2})/2$, $\Phi(r)$ is gravitational potential at
radius $r$ and $M_{\rm c}$ is mass of cold gas. If $f_{\rm c}$ is
the fraction of released gravitational potential energy that is
heating the cold streams, then energy radiated from cold gas
streams is equal to $f_{\rm c}\dot{E}_{\rm grav}$. The rest of the
energy is converted in kinetic energy or is heating the hot
streams of the gas.
In this work we assume that $f_{\rm c}=1$, which gives the upper estimate
of cooling radiation that can contribute to Ly$\alpha$ luminosity.
Observed Ly$\alpha$ luminosity which
originates from a shell between radii $r_{1}$ and $r_{2}$ is equal
to
  \begin{equation}
    \label{eq:8}
 {%
    L_{Ly\alpha}=f_{\alpha}f_{\rm c} \int^{r_{1}}_{r_{2}}
    \dot{E}_{\rm grav}(r)dr,
    }
%  \overfullrule 5pt
%  \mathindent\linewidth\relax
%  \advance\mathindent-259pt
  \end{equation}
where factor $f_{\alpha}$ is the fraction of the energy radiated
from cold gas streams that we see in Ly$\alpha$ line.

\vspace{12pt}

\begin{figure*}
\includegraphics*[width=0.990\textwidth]{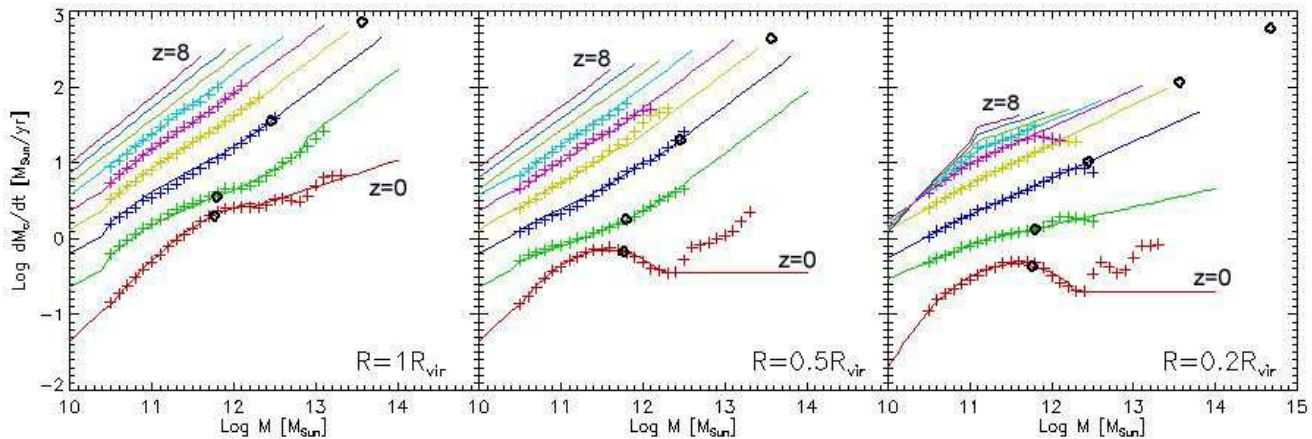}
\caption{Cold gas accretion rates at 1, 0.5, and 0.2 $R_{\rm vir}$
for redshifts $z=0-8$ (from the lowest to the highest values).
Symbols $+$ are FG11 estimates at $z=0,1,2,3,4,5$ (red, green,
blue, yellow, rose and cyan), lines are our fit, and black
diamonds are maximum masses for haloes which could host cold
streams according to
 \citet{DekBir06}. At each redshift we display extrapolated cold gas accretion rates
up to the maximum halo masses from the DM simulation.
}\label{fig:1}
\end{figure*}

\begin{figure}
\begin{center}
  \includegraphics*[width=9cm]{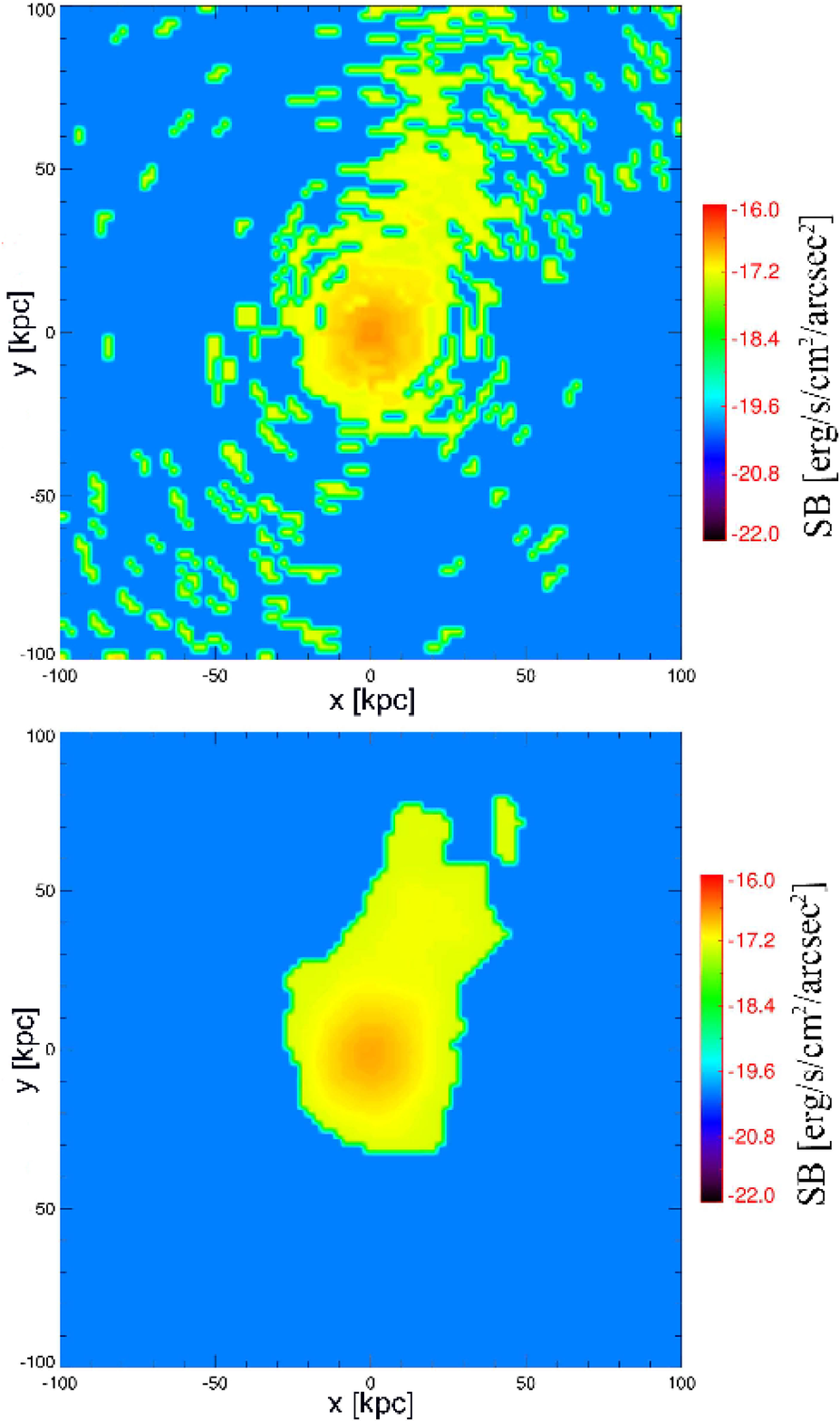}
\end{center}
\caption{Image of one of our LAB at $z=2.3$. Upper image is not
smoothed, and lower image is smoothed with a 2D Gaussian kernel
with FWHM which corresponds to seeing. Different colours
correspond to different surface brightness $\log$
SB.}\label{fig:2}
\end{figure}

%%Now, we describe our model in more detail.
For every halo we calculate Ly$\alpha$ luminosities and areas as follows.
First, we divide each halo into shells between spheres of radii
$L_{\rm soft}\times j$ , $j=0,1,2,...n ,$
where $L_{\rm soft}$ is the
softening length of the DM simulation in (physical) kpc and
$n=L_{\rm soft}/R_{\rm vir}$. The softening length in DM
simulation is almost the same as the pixel size from observations
at redshifts $z \gtrsim 2$. We compute luminosity in every layer by using
eq. \eqref{eq:7} and eq. \eqref{eq:8}. We calculate
gravitational potential, and other quantities as follows:

1) gravitational potential.
For every halo we calculate gravitational potential
from spherically averaging distribution of dark matter particles inside the halo,
following equation given in \citet{Binney}:

  \begin{equation}
    \label{eq:9}
 {
   \Phi(r)=-4\pi G \left[\frac{1}{r}\int^{r}_{0}\rho(r')r'^{2}dr' + \int_{r}^{\infty}\rho(r')r'dr'\right] ,
    }
  \end{equation}
where $\Phi(r)$ is gravitational potential and $\rho(r)$ is mass
density in dark matter at radius $r$.
%%Haloes have various shapes and density distributions, which could differ from \citet{NFW97} profile.

2) cold gas accretion rate $\dot{M}_{\rm c}$ .
We use cold gas accretion rates from the simulation of
FG11 without winds,
primarily because
these authors have provided
$\dot{M}_{\rm c}$
at a few different radii inside a halo.
This is important because $\dot{M}_{\rm c}$
are lower near the halo centres, especially at low redshifts.
FG11 have provided
 median cold gas accretion rates in shells
located at $0.2, 0.5$ and $1 \times R_{\rm vir}$ as a function of
dark matter halo masses at redshifts $z=0,1,2,3,4,5$. These rates
are computed directly using instantaneous velocity vectors of
particles. They include both accretion through mergers and smooth
accretion. Thicknesses of the shells are 0.2, 0.2, 0.1 $R_{\rm
vir}$, respectively.
We estimate $\dot{M_{c}}$ at all relevant masses and redshifts
by interpolating and extrapolating $\dot{M_{c}}$ from FG11 (see Appendix {\S}\ref{sec:X.1}).
In Figure \ref{fig:1} we present
$\dot{M_{\rm c}}$  at 1, 0.5, and 0.2 $R_{\rm vir}$, at different masses and redshifts.

3) $f_{\alpha}$ - fraction of energy that we see in
Ly$\alpha$.
Ly$\alpha$ flux is
diminished by absorption
through the intergalactic medium (IGM),
 particularly at the blue side of the line.
We use effective optical depth of \citet{FG08}
%%   observed Ly$\alpha$ forest at a redshift range $z\sim2-4$, and
%%calculated IGM absorption. In our model we use their fit for
%%continuum-corrected measurement in $\delta z = 0.1$ bins, but
%%which is not corrected for metal absorption
: $\tau_{\rm eff}=0.0018(1+z)^{3.92}$ (their eq. 21). %%At $z=1-4$ it is $\tau_{\rm eff}=0.03-0.99$.
We account for the IGM opacity by multiplying the calculated emissivities
by $f_{\alpha}=0.5+0.5\exp(-\tau_{\rm eff})$ .
Here, we assume that half of the Ly$\alpha$ flux is emitted from the blue side of the line, and that
the whole flux from the red side of the line is transmitted.
If LAB emission originates mainly from cold gas accretion, then
dust opacity within the galaxy would be negligible.

4) Cold gas filaments.
In previous work it was shown that cold gas is placed
along dark matter filaments inside a halo \citep[see e.g.][]{Dek09}.
Considering that we use pure DM simulation without gas,
we further divide each
shell into cells, and assume that
within each shell cold gas is distributed uniformly and only
in cells where dark matter density is larger than the halo's average density in dark matter.
A shell which
is situated between spheres of radii $nL_{\rm soft}$ and
$(n+1)L_{\rm soft}$, $n\geq 1$
is divided into $n+1$ parts, such that their projection onto xy-plane consists of
$n+1$ concentric circles with radii $L_{\rm soft}\times j ,
j=1,2,...,n+1$. Then, each part
whose projection is between circles of radii $jL_{\rm soft}$
and $(j+1)L_{\rm soft}$, $j\geq 1$, is further divided into $2j$
equal cells which enclose the same angles with the halo centre. In
this way, the cells have roughly equal dimensions
and their projection
is straightforward to calculate.
 Then, SB maps are obtained by projecting haloes onto the xy-plane
(see the upper image in Figure \ref{fig:2}).

5) LABs images and influence of the atmosphere (seeing).
The SB images are transferred
into rectangular coordinate system with cells of size $L_{\rm soft}$, for all
redshifts $z \gtrsim 2.3$. Then, at $z \gtrsim 2.3$
 the images are smoothed with a 2D Gaussian kernel of FWHM which
corresponds to the seeing of 1 arcsec, which is approximately
equal to the seeing in LABs surveys.
At $z\sim 1$ softening length in DM simulation is smaller than the 5.3 arcsec spatial resolution
in the corresponding GALEX NUV band observations, thereby at $z\sim 1$ we merge a few cells in order to obtain similar resolution.
  Afterwards, in every halo we find all
  objects which do not contain adjacent cells with common vertices above a chosen
SB threshold. At $z=1,2.3,3.1,4,5,6.6$ we use SB thresholds from
the observations of \citet{Barg12}, \citet{Y09}, \citet{M04},
\citet{Sai06}, and \citet{Ou09}. For every source we calculate its
luminosity by summing the luminosities from all its cells above
the SB threshold (hereafter these luminosities are denoted by $L_{\rm CR,obs}$, while
cooling radiation luminosities from all cells are denoted by $L_{\rm CR,tot}$).
Image of one of our LABs is presented in Figure \ref{fig:2}.
After smoothing, the emission is less fragmented and more similar to the observed LABs,
while, on average, luminosities and areas are smaller.

\subsection{Ly$\alpha$ from star formation}\label{sec:4.2}

Newly formed stars could photoionize surrounding neutral HI gas
and radiate in Ly$\alpha$. Luminosities in Ly$\alpha$ and
H$\alpha$ trace star formation rate (SFR). H$\alpha$ emission
originates mostly from HII regions around young hot stars. SFR is
obtained when this emission is extrapolated to lower mass stars,
by assuming Salpeter initial mass function.
For case B recombination the relation between the emitted luminosity and SFR
is derived from a combination of equation 2 in \citet{Kenn98} and of \citet{Br71}.
The observed luminosity is calculated by multiplying the emitted luminosity by the escape fraction $f_{\rm esc}$,
which denotes the fraction of the emitted luminosity which is not absorbed
by the dust inside a halo or by IGM, and is observed from the Earth. The relation between the
observed luminosity and SFR is:

  \begin{equation}
    \label{eq:10}
 {
    L_{Ly\alpha}= 1.1 \times 10^{42}\times f_{\rm esc} {\rm SFR}.
    }
%  \overfullrule 5pt
%  \mathindent\linewidth\relax
%  \advance\mathindent-259pt
  \end{equation}

We calculate luminosities from SFR as follows:

1) Stellar mass.
For each halo we calculate its stellar mass, using equations given
in \citet{Beh13}. \citet{Beh13} matched observed galaxies to their
host haloes by using dark matter simulations and observed stellar
mass function and star formation rates (SFR), and determined how
stellar mass is related to halo mass and redshift. Their results
are in agreement with observed stellar mass functions and SFRs at
a range of redshifts $z=0-8$. We use their equations (3) and (4):

\begin{eqnarray}
\log_{10}(M_\ast(M_h))&=& \log_{10}(\epsilon M_1) + f\left(\log_{10}\left(\frac{M_h}{M_1}\right)\right) - f(0) \label{e:cosmic_sfh} \nonumber\\
\end{eqnarray}
\begin{eqnarray}
f(x)&=& -\log_{10}(10^{\alpha x} + 1) + \delta
\frac{(\log_{10}(1+\exp(x)))^\gamma}{1+\exp(10^{-x})}.
\end{eqnarray}
\begin{eqnarray}
\label{e:redshift_scaling}
\nu(a) & = & \exp(-4 a^2)\nonumber\\
\log_{10}(M_1) & = & M_{1,0} + (M_{1,a}(a-1) + M_{1,z}z)\nu \nonumber\\
\log_{10}(\epsilon) & = & \epsilon_0 + (\epsilon_a (a-1) + \epsilon_z z)\nu + \epsilon_{a,2} (a-1)\nonumber\\
\alpha & = & \alpha_0 + (\alpha_a (a-1))\nu\nonumber\\
\delta & = &  \delta_0 + (\delta_a (a-1) + \delta_{z} z)\nu\nonumber\\
\gamma & = &  \gamma_0 + (\gamma_a (a-1) + \gamma_{z} z)\nu.
\end{eqnarray}
Here $M_\ast$ is stellar mass, $M_h$ is halo mass, $z$ is
redshift, $a$ is scale factor, and the rest are parameters.
The parameters are taken from \citet{Beh13}.

2) Correcting stellar masses in merger trees.
For some haloes, stellar mass is smaller than in the previous
snapshot, and calculated SFR has a negative value. When two haloes
are merging, some particles become gravitationally unbound,
implying that the halo's virial mass is smaller than in the
previous snapshot. Then, in the following snapshot, merger remnant
forms and new halo virializes. In some cases, this intermediate
snapshot catches the moment of merger when material unbinds and
halo's mass decreases. We correct halo masses for this effect by
interpolating between the snapshots, in those cases where the drop
in halo mass occurs.

3) SFRs.
For a halo $i$ with mass $M$ at redshift $z$ we calculate SFR in the
following manner. We find all progenitors of halo $i$ in previous
snapshot (at redshift $z+{\rm d}z$). Then we calculate stellar
mass M$_*$ of the halo $i$, and subtract stellar masses of progenitor
haloes. This difference is the mass in stars which formed between
two consecutive snapshots. Finally, we divide this mass with the
time interval between $z$ and $z+{\rm d}z$:

  \begin{equation}
    \label{eq:11}
 {%
    SFR(i,z)=\frac{M_{*}(i,z)-\sum_{j=0}^{k}M_{*}(j,z+dz)}{dt}.
    }
%  \overfullrule 5pt
%  \mathindent\linewidth\relax
%  \advance\mathindent-259pt
  \end{equation}
In section {\S}\ref{sec:5.5} we show that
our SFR functions at its bright end
are in agreement with other recent works
\citep[e.g.][]{Tes14}.

4) Luminosities from SFRs.
Ly$\alpha$ luminosities are calculated from eq.
\eqref{eq:10}, where for $f_{\rm esc}$ we have used fit from
\citet{DL13}, $f_{\rm esc}(z)=e^{-4.0 + 0.52z}$.
Their $f_{\rm esc}$ includes IGM opacity.
\citet{DL13} determined $f_{\rm esc}$ by comparing observed SFR functions to
Ly$\alpha$ luminosity functions from observations of LAEs at
redshifts $z=0.35, 3.1, 3.7, 5.7$. They successfully reproduced
observed LAEs luminosity functions with $f_{\rm esc}$ which
depends only on redshift, not on halo mass or SFR, over $\sim 2$
orders of magnitude in the luminosity.
But, when we implemented the same $f_{\rm esc}$ in our model, LFs were not well reproduced at high redshifts,
so we fitted ours (see Figure \ref{fig:5} and $f_{\rm esc 1,2}$ in Figure \ref{fig:4}).

\subsection{Luminosity functions}\label{sec:4.3}

Cumulative luminosity function (LF)\footnote{Note that we use abbreviation LF
for cumulative luminosity function, not for (non-cumulative)
luminosity function.} displays the number density of objects with
luminosities greater than $L$.

However,
volumes with different density could have a range of different
LFs. In order to account for this effect, we chose 1000 random cubical
sub-volumes from the DM simulation box. Each sub-volume has a
volume equal to an observing survey
of LABs at the same redshift. At $z=0.8-1,1.5-2.3,3.1,4-5,6.6$ we
chose surveys of \citet{Keel09} (one half of their total volume,
which would roughly correspond to the volume around one of the
observed clusters), \citet{E11} (which is almost the same as
\citet{Y10} (CDFS)), \citet{M04}, \citet{Sai06}, and \citet{Ou09},
respectively.

%%We define mean LFs as follows.
For each LF we calculate (in log-scale) the area below it on the Figure \ref{fig:4}. These
areas we will call LF-areas. Then, we distribute all LF-areas into
bins, fit a gaussian distribution, and find the mean value. We
define the mean LF as the average of all LFs from the bin in
which the mean LF-area is situated.
%%The mean LFs almost coincide with LFs in the whole simulation box.
LFs which differ by $\pm 2\sigma$ from the mean
are determined from
those two LF-areas which enclose $\pm 2\sigma$ values around the mean LF-area. %% LF-area values

\section{Results}\label{sec:5}

\subsection{Luminosity versus mass}\label{sec:5.1}

\begin{figure*}
\vspace{5mm}  %%10mm
  \includegraphics*[width=0.990\textwidth]{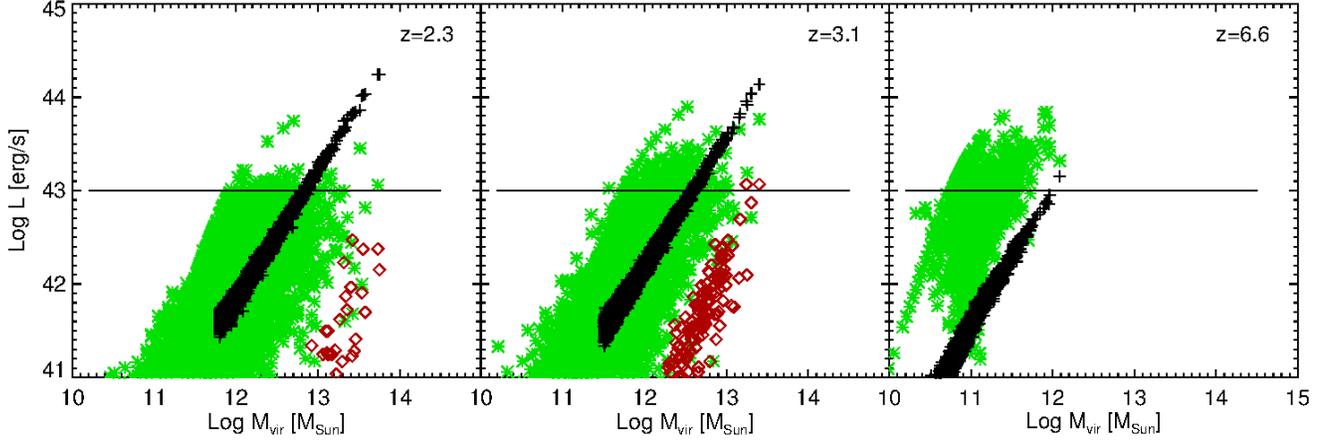}
 \vspace{5mm}
\caption{Luminosity versus mass. Green stars represent
luminosities from star formation, red diamonds are luminosities
from cooling radiation above SB threshold, and black pluses %% blue
are total
luminosities from cooling radiation. Line at $L=10^{43}$ erg s$^{-1}$ represents
a minimum luminosity above which could be detected LABs in most
observations.}\label{fig:3}
\end{figure*}

In Figure \ref{fig:3} we show how the modelled luminosities are
related to halo masses. This includes SF luminosities ($L_{\rm
SF}$); cooling radiation luminosities above a corresponding SB threshold - for
the most luminous source inside a halo ($L_{\rm CR,obs}$; see
{\S}\ref{sec:4.1}); and total cooling radiation luminosities ($L_{\rm CR,tot}$),
which includes total cooling radiation luminosity from all cells.

At $z\sim3$ our relation halo mass - luminosity $L_{\rm CR,tot}$ is
similar to the same relation in
\citet{DL09} model, and in %%to
\citet{FG10} prescription 7.
%%(which is their prescription with accounted self shielding and no luminosity from multiphase
%%star-forming particles).
However, in comparison to \citet{Ros12},
our relation halo mass - luminosity
is steeper, and our luminosities are lower.
  The Figure also shows that
$L_{\rm CR,obs}$ are less than $L_{\rm CR,tot}$ by $\sim$ an order of
magnitude, or more.
This difference is the
most significant at lower redshifts, where haloes are more
extended and have more rarefied gas, and at higher redshifts $z
\gtrsim 5$, where the sources are less luminous.

\subsection{Luminosity functions from cooling radiation}\label{sec:5.2}

\begin{figure*}
  \includegraphics*[width=0.990\textwidth]{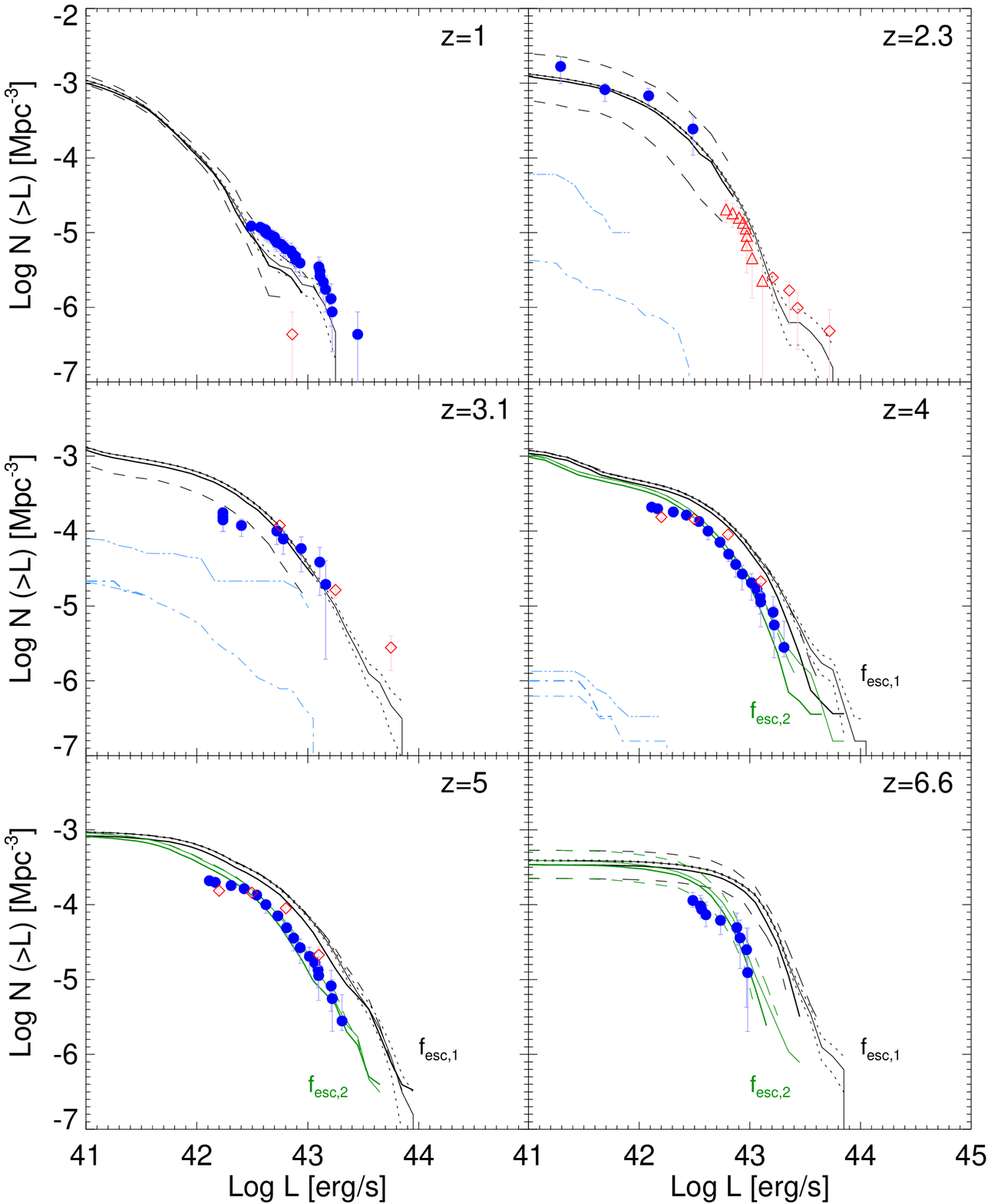}
\caption{LAB and LAE cumulative luminosity functions at a few different redshifts.
 Light blue lines present LFs
from cooling radiation (above SB threshold), black lines are LFs from SF,
 and green lines are LFs from SF but calculated for our $f_{\rm esc}$
($f_{\rm esc}$ are also indicated on the Figure: $f_{\rm esc,1}$ are from \citet{DL13}, while $f_{\rm esc,2}$ are our best fit).
 For LFs from SF (cooling radiation)
full lines (dot-dashed lines) represent LFs in the whole simulation box,
thick full (dot-dashed) lines represent mean LFs,
upper and lower dashed (double-dot-dashed) lines
represent $2\sigma$ value distribution around the mean LFs (see {\S}\ref{sec:4.3}),
and upper and lower dotted lines represent Poisson errors in the simulated LFs.
  Observations are presented with different symbols:
  filled circles are observed LAEs LFs, while
  red symbols are observed LABs LFs.
At $z=2.3$ different LABs observations are denoted with different symbols:
diamonds are \citet{Y09} and triangles are \citet{Y10} (fields).
Observations in fields ($\delta\sim 0$) at other redshifts are summarized in section 3.
Observed LFs are presented with Poisson error bars.}\label{fig:4}
\end{figure*}

In Figure \ref{fig:4} we present cumulative luminosity functions
at different redshifts calculated from:

1) our model with cooling radiation, $L_{\rm
CR}$,

2) our model with star formation, $L_{\rm
SF}$,

3) observations of LAEs and LABs in fields (symbols).

The Figure includes
mean LFs and $\pm 2\sigma$ range from them, and Poisson errors
for both observed and simulated data.
The mean LFs almost coincide with the LFs calculated in the whole simulation box.

At all redshifts LFs calculated from cooling radiation are below the observed
ones. This could be seen particularly at lower ($z\sim1$) and at
higher ($z\sim6$) redshifts. At $z\sim2-4$ our $L_{\rm CR}$ are
too low to explain the observed LABs and LAEs LFs. At $z\sim 1$
and $z\gtrsim5$ we did not obtain any Ly$\alpha$ emission from cooling radiation
above the SB thresholds. Even if for every halo we sum luminosity
from all cells above the SB threshold, our LFs would still be
below the observed ones. Our results show that cooling radiation, as we modelled
it, is insufficient to power most of LABs, at every redshift from
$z=1-6.6$ and at every halo mass.

\subsection{Luminosity functions from star formation}\label{sec:5.3}

At a range of redshifts $z=1-5$
  we obtain a good agreement of our LFs with
observed LABs and LAEs LFs.
At $z\sim 2.3-3.1$
%%- At redshifts $z=2.3$ and $3.1$, where are most LABs found,
observed LABs and LAEs LFs are falling inside $\pm 2\sigma$ from
the mean modelled LFs, eventually with the exception of the most luminous LABs at $z\sim 3$.
At $z=1$ our LF is in a good agreement with the observed LAEs LF,
however somewhat above. This could be explained by additional AGN
activity, as is detected in some of these LAEs.
One LAB detected at $z=1$ is roughly falling inside $\pm2\sigma$.
At $z=4-5$ our LFs are in agreement with the observed LAEs and LABs LFs
%%(taking into account other uncertainties)
, however our values are somewhat above the observations.
\footnote{
Note that LFs at $z=4-5$ show no variance, which is (mostly) because the DM simulation
box ($\sim 6\times 10^{6}$ Mpc$^3$) have similar size as
the observed volume at $z=4-5$ ($\sim 3\times 10^{6}$ Mpc$^3$).
 }

At redshift $z=6.6$
our LF is above the observed LAEs LF,
implying that
our SF luminosities are too high.
This could be explained if the most massive haloes are overabundant or have too high SFRs,
or if $f_{\rm esc}$ is overestimated.
Actually, massive haloes at high redshift are slightly overabundant (\citealt{NM14}), but
we do not expect that this would significantly influence our results.
On the other hand, our SFR functions (SFRFs)
are in agreement with other work (see section {\S}\ref{sec:5.5}),
but note that at high luminosities observations are rare, and results from
various models could differ.
In comparison with Schechter SFRFs
from \citet{Smit12}, which used \citet{DL13},
SFRs from our model are too high in massive haloes at high redshifts.

If the number density of
massive haloes and SFRs are not overestimated significantly, than
$f_{\rm esc}$ should be smaller. In Figure \ref{fig:4} we also show
our LFs, but with $f_{\rm esc}$ for which we obtain the best
agreement with observations (denoted by $f_{\rm esc,2}$).
In this case, for appropriate $f_{\rm esc}$ we obtain a good agreement with the data.
As a function of redshift, our $f_{\rm esc}$ could be fitted with
a function of the same shape as in \citet{DL13}, $f_{\rm
esc}(z)=e^{-a -b z}$, with parameters $a=3.76 , b=-0.38$.  %% Dijkstra: a=4.0, b=-0.52
At $z \geq 4$ we will further use our $f_{\rm esc}$.

\begin{figure}
\begin{center}
  \includegraphics*[width=8cm]{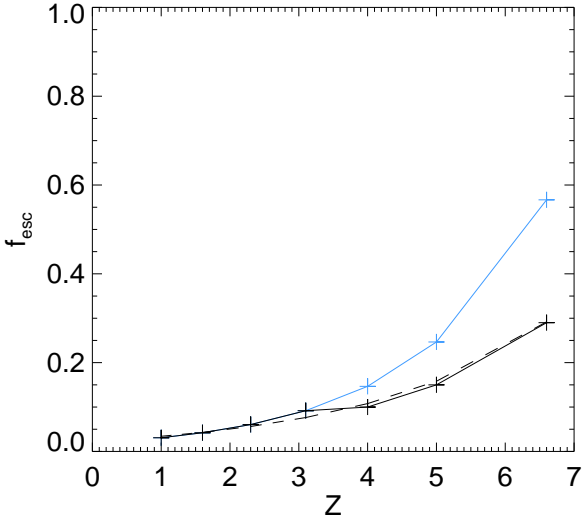}
\end{center} \caption{Escape fraction as a function of redshift.
Blue line is $f_{\rm esc}$ from \citet{DL13}.
Full black line with symbols represents our $f_{\rm esc}$, and
black dashed line represents fit to our $f_{\rm esc}$.
}\label{fig:5}
\end{figure}

In Figure \ref{fig:5} we show our
escape fraction as a function of redshift.
While at $z \lesssim 3$
our $f_{\rm esc}$ are almost the same as in \citet{DL13},
at higher redshifts the $f_{\rm esc}$ are smaller,
but not ruled out from \citet{DL13} with a great significance.
For example, at $z\sim 4$ and $z\sim 6.6$ we found $f_{\rm
esc}\sim0.10$ and $f_{\rm esc}\sim0.29$, respectively, while
\citet{DL13} fit gives $f_{\rm esc}\sim0.15$ ($f_{\rm esc}\sim0.1-0.2$)
and $f_{\rm esc}\sim0.57$ ($f_{\rm esc}\sim0.35-0.92$) (see also discussion in \citet{DL13}).
In addition, note that differences between our model and
\citet{DL13} could influence the calculated $f_{\rm esc}$:
our SFRFs differ from Schechter functions,
we do not have scattering in $f_{\rm esc}$,
we include observed LFs at redshifts higher than $z=5.7$,
and we are not using non-cumulative
luminosity functions. %%(not the non-cumulative ones).
%%\vspace{12pt}

\subsubsection{Influence of overdensity}\label{sec:6.2}

\begin{figure*}
\begin{center}
  \includegraphics*[width=0.990\textwidth]{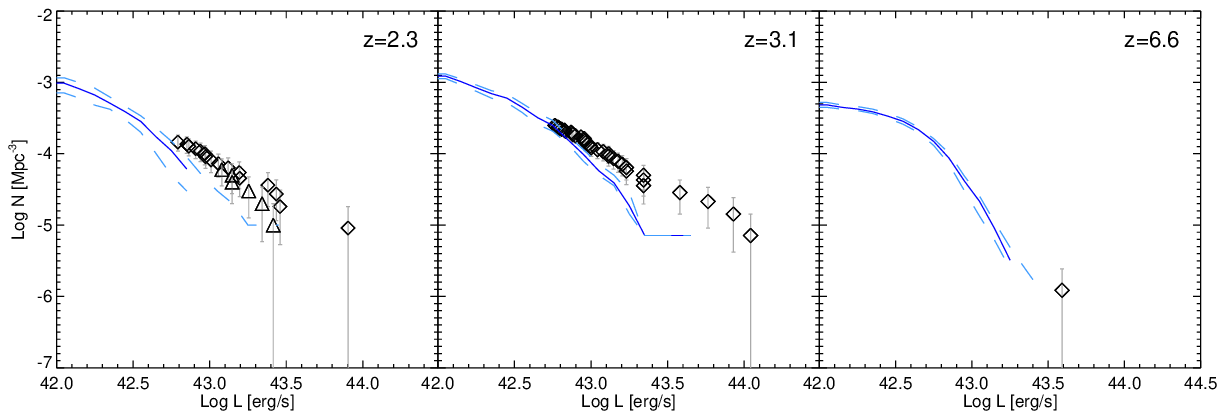}   %% Fig14
\end{center}
\caption{LFs for SF model, as calculated in the most overdense regions,
at redshifts $z=2.3$, 3.1, 6.6. Diamonds represent observations from \citet{Y10} (protocluster; $z=2.3$), \citet{M04} ($z=3.1$) and \citet{Ou09} ($z=6.6$); triangles represent observations from \citet{E11} ($z=2.3$).
}\label{fig:14}
\end{figure*}

Figure \ref{fig:14} represents %% shows
LFs in the most overdense regions at $z=2.3$, 3.1 and 6.6.
We randomly choose 3000 sub-volumes with the same volume as in observations in overdense regions (\cite{Y10} at $z=2.3$, \cite{M04} at $z=3.1$ and \cite{Ou09} at $z=6.6$). We calculate LFs in the most overdense sub-volumes, with dark matter overdensity higher than $\delta_{\rm min}+(\delta_{\rm max}-\delta_{\rm min})\times 0.85$.
 Here, $\delta_{\rm min}$ and $\delta_{\rm max}$ are minimum and maximum density contrast of all sub-volumes.
The Figure shows that
the calculated LFs are roughly in agreement with observations, but somewhat lower at the highest luminosities ($\sim$ half order of magnitude at $z=3.1$).
This could be explained if
the most overdense sub-volumes are less dense than the observed protoclusters.
The maximum density contrasts in dark matter and in the number of LAEs in subvolumes are equal to 0.6, 2, 1.8 ,
and to 0.5, 1.3, 1.4
at $z=6.6,3.1,2.3$, respectively.

\subsection{SFR functions}\label{sec:5.5}

\begin{figure}
\begin{center}
  \includegraphics*[width=0.490\textwidth]{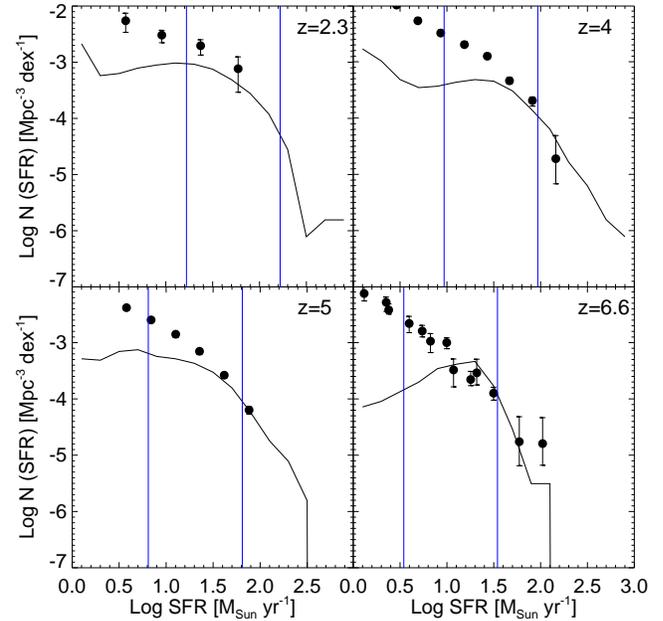}   %%Fig8
\end{center}
\caption{Star formation rate functions from our model at redshifts $z\sim2.3,4,5,6.6$.
Blue vertical lines denote $\log SFR$
which correspond to luminosities $L=10^{42}$ and $10^{43}$ erg
s$^{-1}$. Filled circles are the observations from \citet{Hayes10} (at
$z=2.3$) and from figure 2 in \citet{Smit12} (at $z=4-6.6$).
\citet{Smit12} used results from UV observations in
\citet{Bouwens07,Bouwens11}.}\label{fig:8}
\end{figure}

In order to investigate how accurate our method is, we compare our
star formation rate functions (SFRFs) with other similar works at
$z\sim 2-7$ (Figure \ref{fig:8}).
The Figure shows that
for high SFRs, $\log SFR\gtrsim 1-1.5 M_{\odot}$ yr$^{-1}$, our results are in agreement
with observations (\citet{Hayes10} at $z\sim 2.3$ and \citet{Smit12} at $z\sim 4-6.6$).
These SFRs correspond roughly to luminosities $\log L \gtrsim 42-43$.

However, in comparison with Schechter SFRFs from \citet{Smit12}
(which used \citet{DL13}; not shown in the Figure), SFRs from our
model are higher in massive haloes at high redshifts. For example,
at $z=4$ at $\log SFR\sim2.3$ our SFRF gives $N=-4.8 {\rm Mpc^{-3}
dex^{-1}}$, while \citet{Smit12} gives $N=-5.5 {\rm Mpc^{-3}
dex^{-1}}$.
Similar conclusions hold at $z>4$.
 On the other hand, observations at bright end
are rare, and results from various models
could differ \citep[see e.g.][]{Tes14}. Roughly, our SFRFs at
bright end are in the range between different simulations of
\citet{Tes14}.
  We mention that, as \citet{DL13} discussed, SFRFs
are better described by Saunders functions instead of Schechter
functions \citep{SL12}. These functions are almost identical at
smaller luminosities, but at higher luminosities Saunders
functions decrease more slowly, which is consistent with our
larger number of more luminous haloes.

At small SFRs our SFRFs are too small, due to the
limited resolution for the minimum halo mass $M_{\rm min}$ in DM
simulation.
We could apply our results at $\log SFR \gtrsim 1-1.5 M_{\odot}$
yr$^{-1}$, which corresponds to luminosities $\log L \gtrsim
42-42.5$ erg s$^{-1}$.
Influence of resolution is further discussed in Appendix {\S}\ref{sec:6.4}.

%% MINOR***
%%Note that for masses somewhat larger than $M_{\rm min}$ our SFRFs
%%could be somewhat overestimated because we do not have enough
%%low-mass haloes whose mass we would subtract from a corresponding
%%halo. This could explain slightly larger number of haloes with
%%$\log SFR \approx 1.3 M_{\odot}$ yr$^{-1}$ at $z=6.6$.

\subsection{Relation luminosity - stellar mass}\label{sec:5.6}

\begin{figure}
\begin{center}
  \includegraphics*[width=8cm]{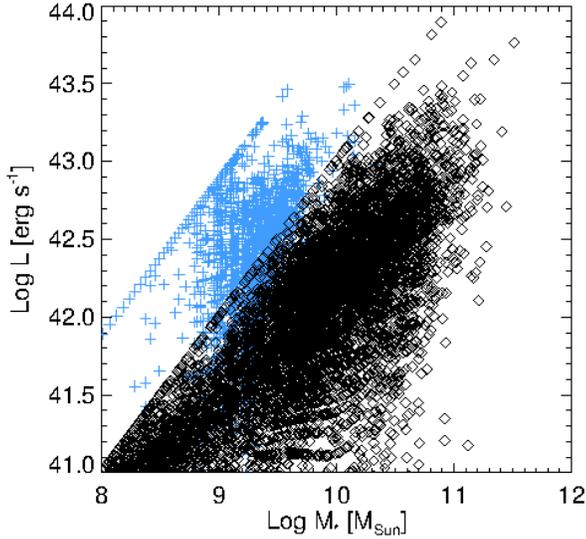}   %%Fig10
\end{center}
\caption{Luminosity as a function of stellar mass, at redshifts $z=3.1$ (black diamonds)
and at $z=6.6$ (light blue pluses).
}\label{fig:10}
\end{figure}

In Figure \ref{fig:10} we present luminosity ($L_{SF}$) as a function of stellar mass.
As expected, haloes with larger masses and situated at higher redshifts show larger luminosities.
  Our results are roughly in agreement with LAB stellar masses estimated in different observations at $z\sim 3-6$.
  LAB stellar masses estimated in observations at $z\sim 3$ are $\sim 1-5 \times 10^{11} M_{\odot}$
  for LABs with luminosity $L\sim 10^{44}$ erg s$^{-1}$ \citep[see][]{Sm08, Martin14} ,  %%Sm08: 4x10^11 (3-5) ; Martin14: 1-4
  and at $z\sim 6.6$ for a LAB with luminosity $L\sim 4\times 10^{43}$ erg s$^{-1}$ the
  estimated stellar mass is $\sim 3.5\times 10^{10} M_{\odot}$ \citep[see][]{Ou09}.
  %% LAEs? Roughly in agreement with Gawiser+09. But large scatter and L range.

\section{Discussion}\label{sec:6}

Sections {\S}\ref{sec:6.1}, {\S}\ref{sec:6.3} and {\S}\ref{sec:6.b} discuss LAB areas, influence of the parametes used, and bias factor of LABs.
Other uncertainties include assumption that particles inside a halo are spherically symmetrically
distributed, lack of radiative transfer calculation \citep[see e.g.][]{FG10}, and lack of
influence of other sources of energy (such as AGN and starburst supernovae).
%%1) substructures and halo mass - gas distribution: We have
%%calculated gravitational potential differences with
%%assumption that particles inside a halo are spherically symmetrically distributed,
%% which is not necessary true. In addition,
%%\citet{Cen12} showed in their work that substructure could boost
%%Ly$\alpha$ emission and extent. With accounted emission from
%%subhaloes our luminosities and areas would be larger.
%%2) propagation of the emitted radiation:
%%we do not have radiative transfer calculation, which
%%could influence calculated luminosities \citep[see e.g.][]{FG10}.
%%3) other galaxies and other sources of energy:
%%In our model we did not include contribution from other sources of
%%energy, such as superwinds driven by starburst supernovae (SN) or
%%photoionization by active galactic nuclei (AGN). It is not clear
%%how much these different sources contribute.
%%Depending how these sources are related to each other, we could obtain the same
%%results but with different contribution from different sources.
%%For example, escape fraction could actually be lower in the most
%%massive haloes, but in these haloes also other sources of energy
%%could be more significant, for example AGN or SN.
%%On the other hand we did not exclude sources which host massive
%%haloes but are not detected as LABs (e.g. radio galaxies; however
%%number density of radio galaxies is $\lesssim 10^{-9}{\rm Mpc^{-3}}$, see \cite{Jarvis01}).
LAB source of energy is discussed in section {\S}\ref{sec:6.6}. %% - {\S}\ref{sec:6.5}.

\subsection{LAB areas in SF model}\label{sec:6.1}

%%In this work a cosmological DM simulation is combined with semi-analytical recipes.
%%Using halo merger trees, for every halo Ly$\alpha$ luminosity is calculated.
%%However, using this method we could not calculate LAB areas.
%%{\color{violet} In addition, if some fraction of the emitted luminosity is below the surface brightness threshold, then the observed luminosities will be lower.}
%%We estimate LABs areas by assuming how luminosity is distributed inside haloes.
By using the method presented in {\S}\ref{sec:4.2}, we could not calculate LAB areas.
We estimate LABs areas by assuming how luminosity is distributed inside haloes.

In previous works the surface brightness profiles of extended Ly$\alpha$ emission are well fitted by
\begin{equation}
  \label{eq:14}
{
SB(r) = C \exp(-r/\beta),
}
\end{equation}
where $C$ and $\beta$ are free parameters, and $r$ is the projected radius \citep[see][]{Steidel11,Momose14,M12}. The parameters ranges are $C=1.4-15.7$ and $\beta=25.2-28.4$ in \citet{Steidel11} for all objects except Ly$\alpha$ absorbers, and $C=0.8-5.3$ and $\beta=5.9-12.6\approx$ const in \citet{Momose14} for LAEs at $z=2.2-6.6$.

Motivated by these results, we assume that in all haloes luminosity is distributed (spherically symmetrically) as in eq. \eqref{eq:14}, with a constant parameter $\beta$.
For a few different $\beta$,
for the calculated total luminosities we determine parameter $C$, and find the luminosities and areas above a surface brightness threshold which correspond to observations.
%%%
In this simplified approach Ly$\alpha$ emission is described by no more than one component \citep[for a more complex model see][]{Wis15}.

\begin{figure*}
\begin{center}
  \includegraphics*[width=0.990\textwidth]{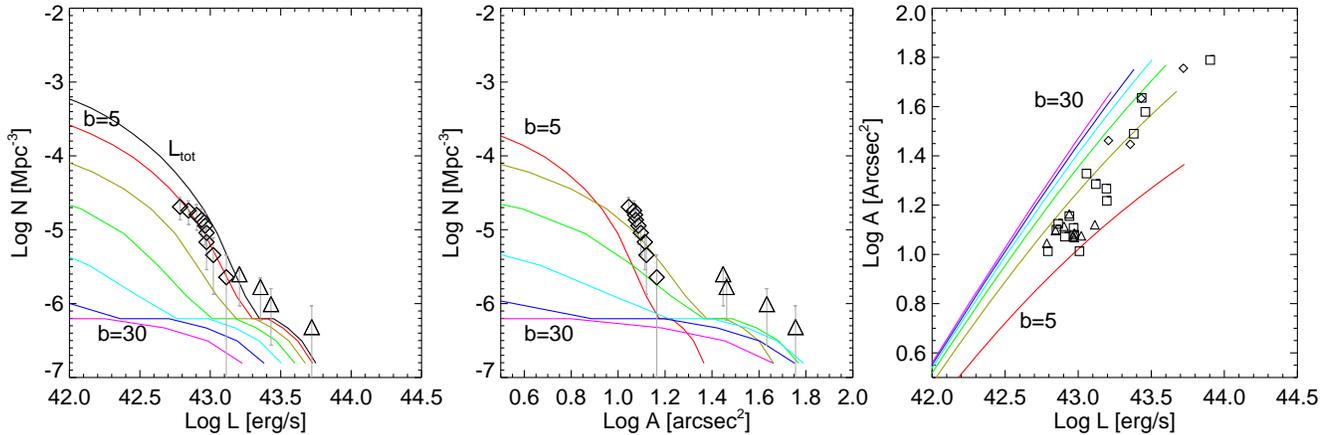}  %%Fig11 ///
\end{center}
\caption{LFs (left), AFs (central) and luminosity-area relation (right)
calculated for SB distribution described with eq. \eqref{eq:14}.
The results are shown for $z=2.3$ and SB threshold as in \citet{Y10}.
Different lines correspond to different parameter $\beta$,
in the right (left and central) figure from bottom to upper (upper to botom): $\beta=$ 5,10,15,20,25,30.
Black lines (denoted with $L_{\rm tot}$) correspond to total luminosities.
Symbols represent different observations: diamonds represent results for fields in \citet{Y10},
triangles are for \citet{Y09}, and squares are for protocluster in \citet{Y10}.
}\label{fig:11}
\end{figure*}

In Figure \ref{fig:11}
we show LFs, cumulative area functions (AFs; defined in the same way as LFs) and the relation luminosity -- area, calculated for a few different parameters $\beta$.
The results are shown for
 observations of \citet{Y10} (fields) and \citet{Y09}, at redshift $z=2.3$ and for surface brightness (SB) threshold $SB_{\rm thr}=5.5 \times 10^{-18}$ erg s$^{-1}$ cm$^{-2}$ arcsec$^{-2}$.
From the Figure \ref{fig:11} one can see that for $\beta\sim 5-10$ the calculated LFs, AFs and the relation between luminosities and areas are roughly in agreement with observations. The calculated LFs for $\beta\sim 5$ are almost the same as LFs for total luminosities.
As $\beta$ increases, luminosities and areas are smaller, and areas increase faster with luminosities.

\begin{figure}
\begin{center}
  \includegraphics*[width=8cm]{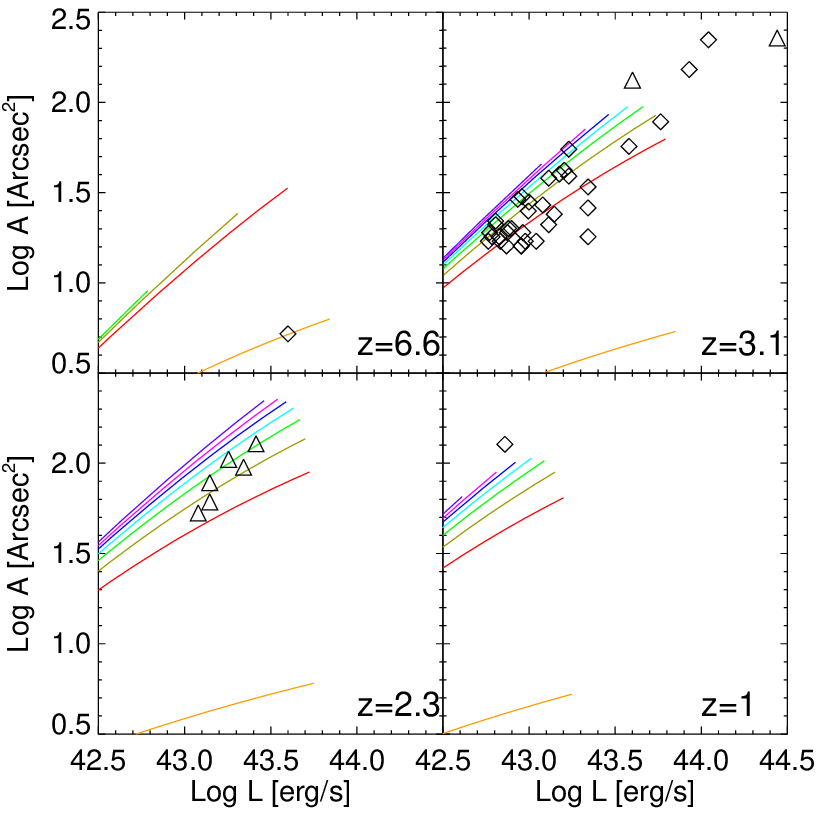}   %%%[width=0.490\textwidth]{Fig13.eps}
\end{center}
\caption{Luminosity-area relation calculated for a few different observations, from the upper left to the bottom right:
\citet{Ou09} ($z=6.6$),
\citet{M04} ($z=3.1$; triangles represent two LABs observed by \citealt{M09}), \citet{E11} ($z=2.3$), \citet{Barg12} ($z=1$).
Observations are represented by symbols.
Different lines correspond to different parameter $\beta$, from bottom to upper: $\beta=$ 1.4,5,10,15,20,25,30,42 (at some $z$ only the lines for the lowest $\beta$ are shown).
}\label{fig:13}
\end{figure}

In Figure \ref{fig:13} we show the luminosity-area relation for a few different observations, at different redshifts. At $z\sim2-3$ we obtain agreement with observations for $\beta\sim 5-10$.
For LABs observed at $z=2.3$ by \citet{E11} above a lower SB threshold, $SB_{\rm thr}=1.5 \times 10^{-18}$ erg s$^{-1}$ cm$^{-2}$ arcsec$^{-2}$, we find $\beta\sim15$.
However, at $z=1$ and at $z=6.6$ we could obtain agreement with observations only if $\beta\sim 1.4$ and $\beta\gtrsim 42$, respectively.
Our results at $z\sim 2.3$ are in agreement with \citet{Momose14}, who obtained $\beta\sim 5-10$ at $z=2.2-6.6$.
However, our results are not in agreement with \citet{Steidel11}, who obtained $\beta=27.6$ for LABs.
We speculate that
this could be explained if the geometry of LABs influences significantly on the determined parameter $\beta$.
It is observed that LABs have asymmetric shapes \citep[see e.g.][]{M11}.
If stacked images of LABs have more shallow SB profiles and larger $\beta$, then LABs with asymmetric shapes could show larger luminosities and smaller areas (than in the case when they have symmetrical shapes), which apparently corresponds to lower $\beta$.
On the other hand, luminosity -- area relation for two LABs observed at $z=1$ and $z=6.6$
indicates that, assuming that the presented model well describes LABs,
the parameter $\beta$ could increase with redshift,
which is not in agreement with \citet{Momose14}.

\subsection{LFs for a range of parameters}\label{sec:6.3}

In this section we present LFs at a few redshifts $z$, calculated for different parameters,
from SF and from cooling radiation models.

\begin{figure}
\begin{center}
  \includegraphics*[width=8cm]{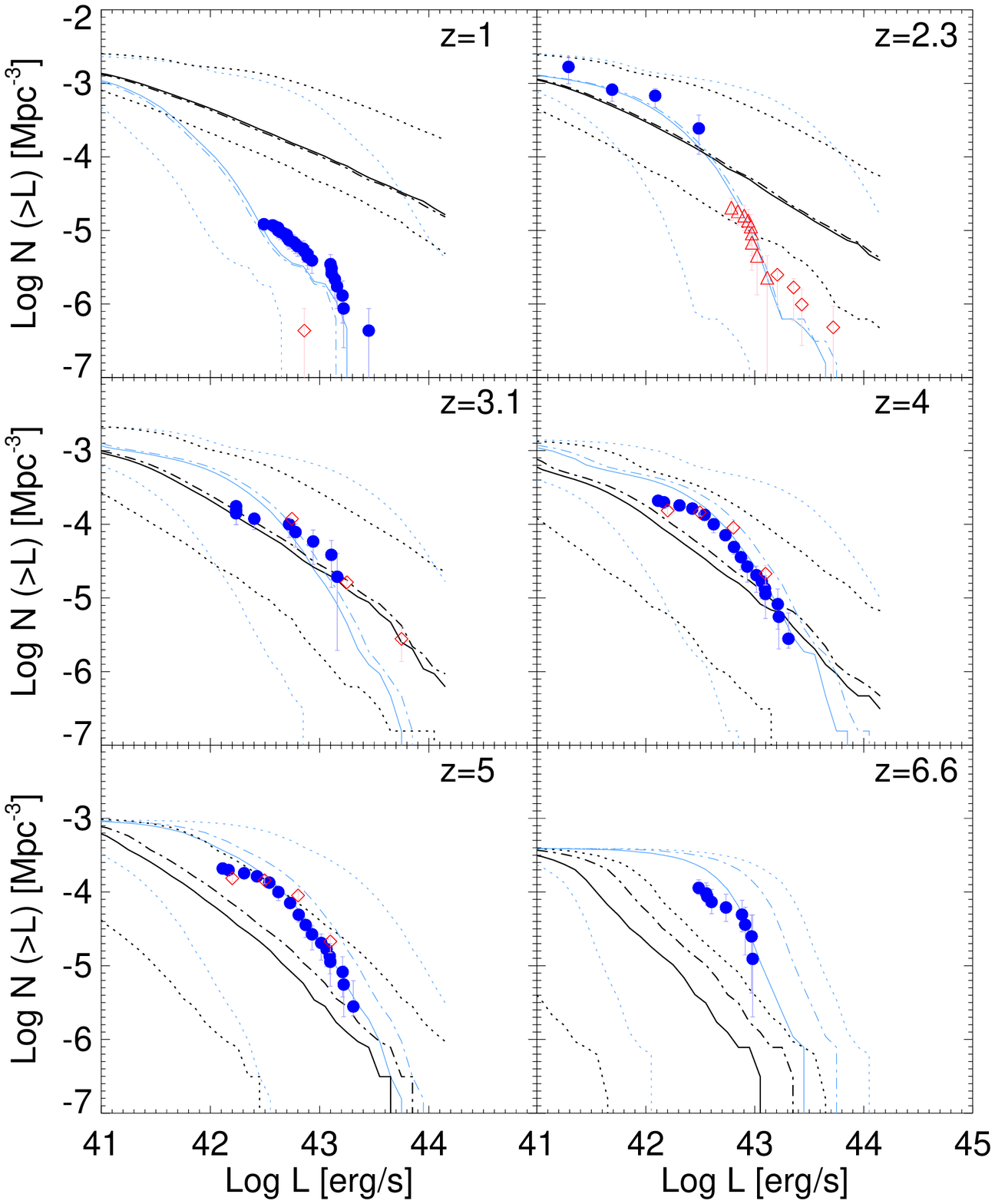}  %%Fig15
\end{center}
\caption{LFs calculated from SF at a few different redshifts.
The notation is the same as in the Figure \ref{fig:4}, but the results are shown for
different stellar masses (black thick lines are for \citet{Mos13}, and light blue lines are for \citet{Beh13})
and for different escape fractions (full lines are for $f_{\rm esc}$ as calculated in \citet{DL13}, dot-dashed lines are for $f_{\rm esc}$ as calculated in our fit,
dotted lines correspond to fixed $f_{\rm esc}=0.01$ and $f_{\rm esc}=1$).
}\label{fig:15}
\end{figure}

Luminosities from star formation.
Escape fractions and relation between stellar masses and dark matter masses are not well determined
\citep[see e.g.][]{Kravtsov14,SW14,DL13}.
For example, \citet{Kravtsov14} showed that at high masses \citet{Beh13}
underestimated stellar masses because they used observations which
did not account for the outer SB profiles of their galaxies.
%%\citet{SW14} and \citet{Kravtsov14} showed that at low and at high masses, respectively, stellar masses \citet{Beh13}}
%%for every halo we determined its stellar mass from relation
%%derived by \citet{Beh13}. However, such relation could be
%%inaccurate in some halo masses if not all haloes contain a galaxy.
%%\citet{SW14} showed that the method of abundance matching based on
%%dark matter only simulations is not accurate at the low mass end
%%because not all low mass haloes host a visible galaxy, and due to
%%presence of baryons haloes grow at a lower rate. In addition,
%%\citet{Kravtsov14} showed that at high masses \citet{Beh13}
%%underestimated stellar masses because they used observations which
%%did not account for the outer SB profiles of their galaxies.
%% On the other hand, since \citet{Beh13} obtained a good agreement with observed
%%stellar mass functions and (cosmic and specific) SFRs, we do not
%%expect that our calculated SFRs would differ significantly from observations.

In Figure \ref{fig:15} we show LFs calculated from SF, but for    %% the Figure 15
different stellar masses (from \citet{Beh13} and from \citet{Mos13}, their eq. (2),(11)-(14))
and escape fractions (as in \citet{DL13}, as in our fit, and for fixed values of 1 and 0.01).
For stellar masses in \citet{Mos13} we use their eq. (2) for both central and satellite galaxies
(note that this could influence on the results to some extent).
One can see that in order to obtain agreement of
LFs calculated using stellar masses from \citet{Mos13} with observations,
the $f_{\rm esc}$ should be higher at luminosities $L\sim 10^{42} - 10^{43}$ erg s$^{-1}$
at redshifts $z\gtrsim 4$, and
at luminosities $L\gtrsim 10^{43}$ erg s$^{-1}$
and redshifts $z\lesssim 2.3$ it should be $f_{\rm esc}<0.01$.
%%%***************************************************************************************

\vspace{10pt}

Luminosities from cold gas accretion rates.
We calculate luminosities and areas for the first 100 haloes (within the largest friends-of-friends masses), for
different combinations of parameters: %%as presented aboove

1) $\dot{M_{\rm c}}$ as calculated in \citet{FG11} and in \citet{Voort11}
(see Appendix {\S}\ref{sec:X.2}).
In differrent work different cold gas accretion rates ($\dot{M_{\rm c}}$) are obtained
\citep[see e.g.][]{FG11,FG10,Voort11,Ros12,Nel13,Benson}.
   For example, at $z\sim 3$ at $\log M = 12-12.5
M_{\odot}$ cold gas accretion rates in \citet{Voort11} are by
about order of magnitude higher than in FG11.
On the other hand, \citet{Nel13}
obtained smaller cold gas accretion rates onto
massive galaxies at $z = 2$ by a factor of up to two or more (see their fig. 3).

2) different slope of the extrapolation of the $\dot{M_{\rm c}}$ at high masses ($\pm 0.1 \times (\log M - 12)$)

3) different distribution of the $\dot{M_{\rm c}}$ at a fixed radius $r$: $f=1-20$, $g=0-10$.
  Simulations showed that cold gas is distributed along filaments of dark matter, where density is a few times higher than the average density inside a halo \citep[see][]{Dek09}.
We assume that the filaments are found in cells where the density is at least $f$ times higher at the virial radius, at least $g$ times higher at the halo center, and at radius $r$ at least $f+(g-f)\times(1-r/R_{\rm vir})$ higher; and we assume that cold gas is homogeneously distributed at each $r$.

\begin{figure*}
\begin{center}
  \includegraphics*[width=0.990\textwidth]{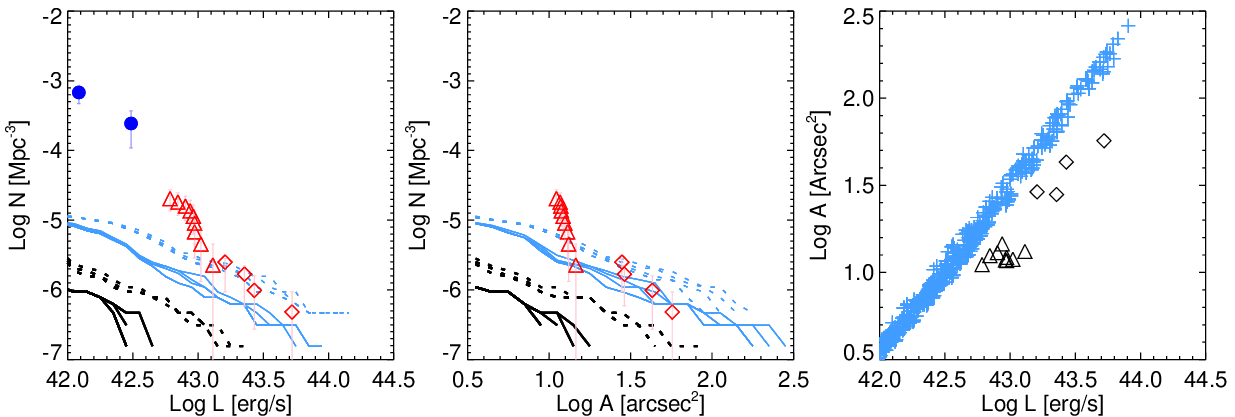} %%{xxx_A9.eps}   %%Fig16
\end{center}
\caption{LFs (left), AFs (central) and luminosity-area relation (right)
calculated for cooling radiation model.
The results are shown for $z=2.3$ and SB threshold as in \citet{Y10}.
For LFs and AFs thick black and light blue lines represent results for $\dot{M_{\rm c}}$ as calculated in \citet{FG11} and in \citet{Voort11}, respectively.
Full and dashed lines represent lower and higher slopes of the  extrapolation of the $\dot{M_{\rm c}}$ at high masses, respectively. For the luminosity-area relation, light blue crosses represent the results calculated for all these cases.
 Symbols represent observations of LAEs (blue circles), LABs in \citet{Y09} (diamonds) and LABs in \citet{Y10} (fields; triangles).
}\label{fig:16}
\end{figure*}

\begin{figure}
\begin{center}
  \includegraphics*[width=8cm]{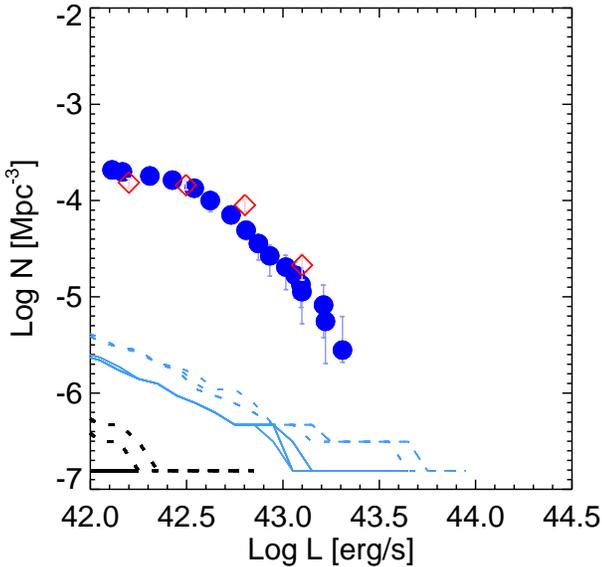}  %%{xxx_CRparr.eps}  %%0.990\textwidth   %%Fig17
\end{center}
\caption{The same as in the Figure \ref{fig:16}, but for luminosities of \citet{Sai06} at $z\sim 3-5$.
}\label{fig:17}
\end{figure}

From Figure \ref{fig:16} one can see that using $\dot{M_{\rm c}}$ as calculated in \citet{FG11}   %%the Figure
LFs and AFs are below the observations.
However, by using $\dot{M_{\rm c}}$ as calculated in \citet{Voort11}, one can obtain agreement with observed LFs or AFs
at $z=2.3$ for appropriate parameters, but the luminosity -- area relation is above the observations.
On the other hand, Figure \ref{fig:17} shows that
at $z\sim 4$ the LFs are below the observations in all cases.

%%\subsection{Bias factor of LABs}\label{sec:6.b}  %%% (as submitted to MNRAS)
\subsection{Overdense regions and fields}\label{sec:6.b}

Observations showed that in overdense regions, usually traced by
LAE number counts, the number
density of LABs is also higher \citep[e.g.][]{Y10}.
This implies that LAB cumulative luminosity functions (LFs) and number densities from different observations could not be directly compared, if the observed volumes have different densities. However, LABs are rare and are in many cases detected in protoclusters with different densities.

In this section it is described how the observed number densities
of LABs from protoclusters with diverse matter densities could be compared to
LABs detected in fields.
In this way, for each observation of LABs it could be predicted what number of LAB would be detected in a volume of
the same dimensions, but with average density (i.e. in the field). In this manner, LABs LFs could be compared between a few different surveys, and it could be roughly estimated how LAB number density (or LF) changes with redshift.
For the discussion on influence of overdensity on calculated LAB LFs see also \citet{Y10} and appendix in \citet{DL09}; and for estimates of evolution of LAB number density with redshift see e.g. \citet{Keel09}.

Density contrast of LABs in some specific volume is defined as $\delta_{\rm LAB}=\frac{n_{\rm LAB}-\bar{n}_{\rm LAB}}{\bar{n}_{\rm LAB}}$,
where $n_{\rm LAB}$ is the number density of LABs in the volume,
and $\bar{n}_{\rm LAB}$ is the mean number density of LABs.
Bias defines how the number density of some objects traces the distribution of the matter density.
Bias of LABs ($b_{\rm LAB}$) is defined as
  \begin{equation}
    \label{eq:1}
 {
    \delta_{\rm LAB}=b_{\rm LAB} \delta_{\rm m} ,
    }
  \end{equation}
where $\delta_{\rm LAB}$ is the density contrast of LABs,
and $\delta_{\rm m}$ is the matter density contrast.
Density contrast and bias of LAEs ($\delta_{\rm LAE}$ and $b_{\rm LAE}$) are defined in the same way.

 We use the relative bias parameter between LABs and LAEs, $b_{1}$, which is defined as:
  \begin{equation}
    \label{eq:2}
 {%
    b_{1}=\frac{\delta_{\rm LAB}}{\delta_{\rm LAE}}=\frac{b_{\rm LAB}}{b_{\rm LAE}} .
    }
  \end{equation}
Equations \eqref{eq:1} and \eqref{eq:2} show that, if in a volume (protocluster) observed at redshift $z$
the number density of LABs is $n_{\rm LAB}$ and the LAEs density contrast is $\delta_{\rm LAE}$,
then the number density of LABs in an average volume at $z$ is
    \begin{equation}
    \label{eq:3}
 {%
    \bar{n}_{\rm LAB}=n_{\rm LAB}/(b_{1}\delta_{\rm LAE}+1) .
    }
  \end{equation}

For the most luminous LABs
we derive $b_{1}$ using values of bias factor of
LABs ($b_{\rm LAB}$) and LAEs ($b_{\rm LAE}$) which are calculated
from observations.
\citet{Y10} have determined bias factor for 6 most luminous and
largest LABs discovered in their survey at $z=2.3$ (with $L
\gtrsim 1.5 \times 10^{43} $ erg s$^{-1}$, $A>16 {\rm
arcsec^{2}}$). They found that $b_{\rm LAB}\sim7$.
%%From the observed subfields of \citet{Ou08} and
%%\citet{Shi09}, \citet{Y10} have also determined variance in
%%$n_{\rm LAE}$ at redshifts $z=3-5$: $\sigma_{v}\sim0.3$. This
%%corresponds to the bias of $b_{\rm LAE}\sim2$.
%%If we assume that $b_{\rm LAE}$ is identical at $z=2.3$ as at $z\sim 3-5$,
\citet{Gua10} have determined bias factor for LAEs at $z\sim2.1\approx2.3$, and found that $b_{\rm LAE}\sim1.8$
\citep[for other references see also][]{Ou10}.
%% b = 1.8 pm 0.3
From eq. \eqref{eq:3}, $b_{\rm LAB}$ and $b_{\rm LAE}$
we obtain
    \begin{equation}
    \label{eq:6}
 {%
    b_{1}\sim3.9  ,   %%3.5  ,
    }
  \end{equation}

%%%%%%%%%%%%%%%%%%%
Now we roughly estimate bias for less luminous and less
large LABs in \citet{Y10} survey, with $L<10^{43}$ erg s$^{-1}$
and $A<16 {\rm arcsec^{2}}$.
We denote this bias factor by $b_{2}$, and mention that it is defined in the same way as $b_{1}$.
%%In order to do this,
First, we estimate
variance using standard equation (\citealt{Peebles}, \S 36):
  \begin{equation}
    \label{eq:12}
 {%
    \sigma_{v}^{2}= \frac{\langle N^{2} \rangle - {\langle N \rangle}^{2}}{{\langle N \rangle}^{2}} - \frac{1}{\langle N
    \rangle},
    }
%  \overfullrule 5pt
%  \mathindent\linewidth\relax
%  \advance\mathindent-259pt
  \end{equation}
and apply it to 4 fields which \citet{Y10} observed.
This equation was also used by \citet{Y10}, but for the most luminous LABs.
We derive the bias factor for these less luminous LABs in the same way as
for the most luminous LABs, and obtain
  \begin{equation}
    \label{eq:13}
 {%
    b_{2}=1.4,   %%1.26,
    }
%  \overfullrule 5pt
%  \mathindent\linewidth\relax
%  \advance\mathindent-259pt
  \end{equation}

We will proceed further with the assumption that the the relative bias parameter between LABs and LAEs %%factor $b_{1}$
is constant with redshift
and that for luminosities $L\gtrsim 1\times 10^{3}$ erg s$^{-1}$ it is equal to
$b_{1}=3.9$, while for luminosities $L\lesssim 1.5\times 10^{3}$ erg s$^{-1}$ it is $b_{2}=1.4$.
%%$b_{1}=3.5$, $b_{2}=1.26$.

\vspace{12pt}

\begin{figure*}
  \includegraphics*[width=0.990\textwidth]{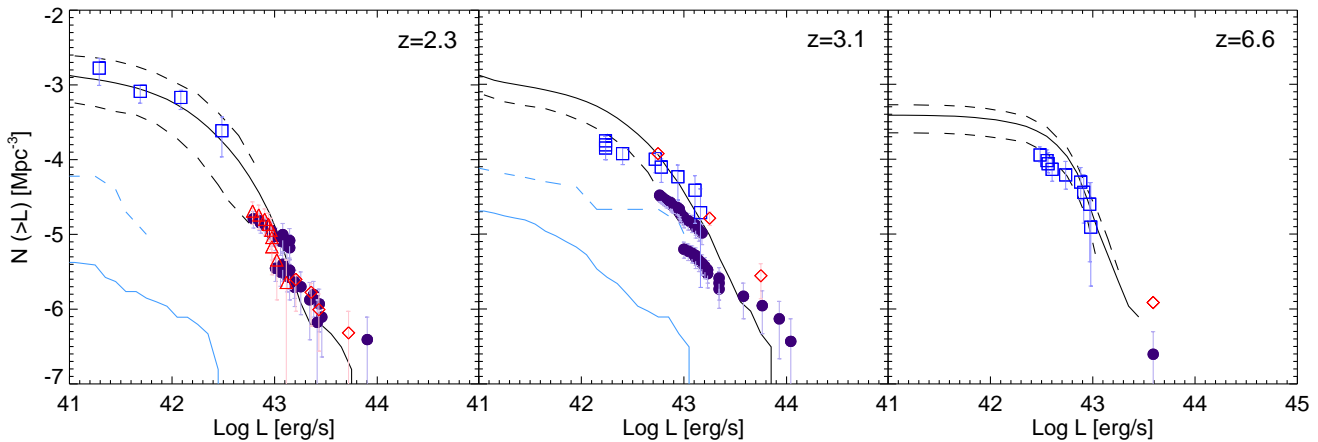}
\caption{LAB and LAE cumulative luminosity functions at $z=2.3$, 3.1 and 6.6.
 Light blue lines present LFs from cooling radiation, while black lines are LFs from SF.
Dashed lines represent $2\sigma$ value distribution around the mean LFs.
Symbols represent observed LAB and LAE LFs.
Observed LAE LFs are represented with blue empty squares, and observed LAB LFs in fields at $z\sim2-3$
are represented with red empty triangles and diamonds.
Observations in overdensities are represented with filled violet circles, and include:
\citet{Y10} (protocluster; $z=2.3$), \citet{M04} ($z=3.1$), \citet{E11} ($z=2.3$) and \citet{Ou09} ($z=6.6$).
For comparison, red empty diamond at $z=6.6$ represents data uncorrected for overdensity for \citet{Ou09} LAB.
}\label{fig:bias}
\end{figure*}

Figure \ref{fig:bias} represents LFs at $z=2.3$, 3.1 and 6.6, but which also include
LFs from observations in the most overdense regions which are corrected for density contrast
by using eq. 12.
The Figure shows that the corrected LFs in overdensities are in agreement with observed LFs
from fields. These LFs are also in agreement with LFs from our SF model.

%%%%%%%%%%%%%%%%%%%%%%%%%%%%%%%%%%%%%%%%%%%%%%%%%%%%%%%%%%%%%%%%%%%%%%%%%%%%%%%%%%%%%%%%%%%%%%%%%%%%%%%%%%%%%%%%%%%

\subsection{LAB source of energy}\label{sec:6.6} %% and comparison with some observations}\label{sec:6.6}

%%In agreement with observations, we have found that LABs are situated in
%%the most massive haloes and in overdense regions at high
%%redshifts. LAEs have luminosities of $\sim 10^{42}-10^{43}$ erg
%%s$^{-1}$, which in our SF model corresponds to halo masses of
%%$\sim 10^{11}-10^{12} M_{\odot}$ at $z\sim 3$. This is in
%%agreement with observations of \citet{Gaw07e}, but in comparison
%%to \citet{Ou10} (their table 5) our LAE host halo masses are higher.

For appropriate $f_{\rm esc}$ our LAB and LAE LFs for SF model
are in agreement with observations at $z\sim1-6.6$,
which is consistent with work in which SF might be the dominant source of energy in majority of LABs
\citep[e.g.][]{Cen12,Col11,Steidel11}.
However, there is also evidence that SF and AGN could be insufficient to
explain luminosities in some LABs (e.g. \citealt{N06}, but see also \citealt{Pr15}). %%\citep[but see also][]{Pr15} .
If this is true, this might indicate that we underestimated cooling radiation luminosities in some cases.
If we would take into account duty cycle
and variations of cold gas accretion rate in different haloes,
then it could be possible to obtain larger cooling radiation luminosities in
some, but not the majority, of the haloes. %% with shorter duty cycles
%%As \citet{Col11} found, majority of LABs could be powered by outflows or photoionization
%% from intense SF, AGNs could be significant in others, and cooling radiation
%% could have important contribution in minority of LABs which could have shorter duty cycles.
In addition, physics of cooling radiation is still not well understood (see
discussion in \citealt{FG10}). For example, it is not well explored
if the infalling streams of the cold gas maintain constant
velocity and continuously radiate their gravitational energy, or
if the streams freely fall into the haloes and release their
energy only when they encounter the central galaxy through a
shock.
%%It is also possible that the infalling gas interacts with
%%substructure in the halo through hydrodynamic instabilities below
%%the resolution of simulations.
Also, for different cold gas accretion rates our results could change.

As is found in some observations, LABs could be complex phenomena
which contain multiple galaxies and fragments of gas, which are
powered by multiple sources of energy with different contribution
in diverse LABs \citep[e.g.][]{Colbert08,Prescott11,Col11,Francis12,Martin14}.
It is also expected that different sources of energy are related to each
other and that photoionization models require cold spatially
extended gas in LABs' host haloes (see e.g. discussion in
\citealt{DL09}).

\section{Conclusions}\label{sec:7}

In this work we have modelled LAB emission from cooling radiation
from the intergalactic gas accreting onto galaxies and from
star formation (SF). We have used a dark matter (DM) simulation,
to which we applied
semi-analytic recipes. These recipes include
cold gas accretion rates from a hydrodynamical simulation, stellar
masses from matching of DM haloes to observed galaxies, escape
fraction of Ly$\alpha$ photons from comparison of observed SFR
functions to Ly$\alpha$ luminosity functions, and intergalactic
opacity from observations of Ly$\alpha$ forest. The advantage of
our model is that we have a large volume and massive haloes at the
same time.
%%We compared our results with observations of LABs (both in fields and in overdensities) and
%%LAEs at a range of redshifts $z\sim 1-6.6$. We found that our cumulative
%%luminosity functions (LFs) from SF are in agreement with the observed
%%LAEs and LABs LFs in fields at all redshifts, and roughly in agreement with LABs LFs in protoclusters.
%%For appropriate choice of free parameters we obtain LABs areas in agreement with observations.
%%However, our results depend on the parameters chosen, such as escape fraction, and on
%%details in calculating how the Ly$\alpha$ emission is distributed.
%%and in overdense regions  Our results are in reasonable agreement with other work.
%%In contrast, our LFs from cooling radiation
%%are too small at all redshifts.
%%However, we also discuss uncertainties in the model which could influence the obtained
%%results.

Here we summarize our main conclusions:

1) We found that luminosities from cooling radiation are too small to explain observed LAB LFs, if we use cold gas accretion rates ($\dot{M}_{\rm c}$) from \citet{FG11}.
However, if we use $\dot{M}_{\rm c}$ from \citet{Voort11}, then our cumulative
luminosity functions (LFs) are in agreement with observations at $z\sim2.3$ for the most luminous LABs ($L > 10^{43}$ erg s$^{-1}$), but at $z\sim4$ our luminosities are still below observations.
Our cooling radiation luminosities are almost independent on uncertainties in cold gas distribution, but to some extent depend on the extrapolation slope in $\dot{M}_{\rm c}$.

2) If we use escape fraction ($f_{\rm esc}$) from \citet{DL13}, then our LFs from SF model are in agreement with observed LABs and LAEs LFs, but only at $z\sim1-3$. However, when we used our $f_{\rm esc}$, that is $f_{\rm esc} (z)$ of the same shape as in \citet{DL13} but for which we found the best fitting parameters, then we found that our modelled LFs are in agreement with all observations of LAEs and LABs in fields at $z\sim1-6.6$.
For LABs in protoclusters our results are also in agreement with observations, but at high luminosities observed LABs might be somewhat more luminous. We note that our results might be dependent on the stellar masses used, and that they are in agreement for \citet{Beh13} stellar masses used.

3) For SF model, we could reproduce luminosity -- area relation of LABs at $z\sim2-3$ for assumed exponential distribution of light inside haloes (eq. \eqref{eq:14}), with a slope parameter $\beta\sim5-10$.
On the other hand, LAB areas at $z\sim 1$ and at $z\sim 6.6$ could be explained if parameter $\beta$ has lower or higher value than at $z\sim2-3$.

4) Our results indicate that majority of LABs and LAEs at a range of redshifts $z\sim1-6.6$ might be powered mainly by SF.
However, we note that there exist other uncertainties in model, and that our results are dependent on parameters used.

\vspace{12pt}

{\bf Acknowledgements} This work was supported by the Ministry of
Education, Science and Technological Development of the Republic
of Serbia through project no. 176001 ``Astrophysical spectroscopy
of extragalactic objects'' and project no. 176021 ``Visible and
Invisible Matter in Nearby Galaxies: Theory and Observations''.
We thank referee whose comments significantly improved the content of
this paper, Milan Bogosavljevi{\' c} for initial idea and useful consultation during the work,
Anne Verhamme for helpful comments on the manuscript.
We also thank Masami Ouchi for helpful comments and discussion, and
Mark Dijkstra, Luka Popovi{\' c} and Du{\v s}an Kere{\v s} for useful discussion.
Numerical results for the DM simulation were obtained on PARADOX cluster at the Scientific Computing Laboratory
of the Institute of Physics Belgrade, supported in part by the
national research project ON171017, funded by the Serbian Ministry
of Education, Science and Technological Development.
The Millennium-II Simulation databases used in this paper and the web application providing online access
to them were constructed as part of the activities of the German Astrophysical Virtual Observatory (GAVO).

\appendix
\section{Appendix}\label{sec:X}

\subsection{Fit for cold gas accretion rates}\label{sec:X.1}

%%FG11 provided estimates of average cold gas accretion rates
%%$\dot{M_{\rm c}}$ at radii $r=1,0.5,0.2$ $R_{\rm vir}$ at a range
%%of redshifts.
In order to estimate $\dot{M_{\rm c}}$
at a given radius $r$ inside a halo ($r=1,0.5,0.2$ $R_{\rm vir}$), as a function of halo mass and redshift,
we interpolate and extrapolate
FG11 values for a few groups of ranges of halo masses and
redshifts (Table \ref{tab:4}).
Between these radii we linearly interpolate $\dot{M_{\rm c}}$.
At high masses, we ignore
an increase of $\dot{M_{\rm c}}$ at $z=0$ for $r=0.5$ and 0.2 $R_{\rm vir}$
and a decrease of $\dot{M_{\rm c}}$ at $r=0.2 R_{\rm vir}$ in FG11
, since there is a relatively small number of the most massive haloes.
From Figure
\ref{fig:1} we see that this fit well describes the
$\dot{M}_{\rm c}$ from \citet{FG11}.

\begin{figure}
\begin{center}
  \includegraphics*[width=8cm]{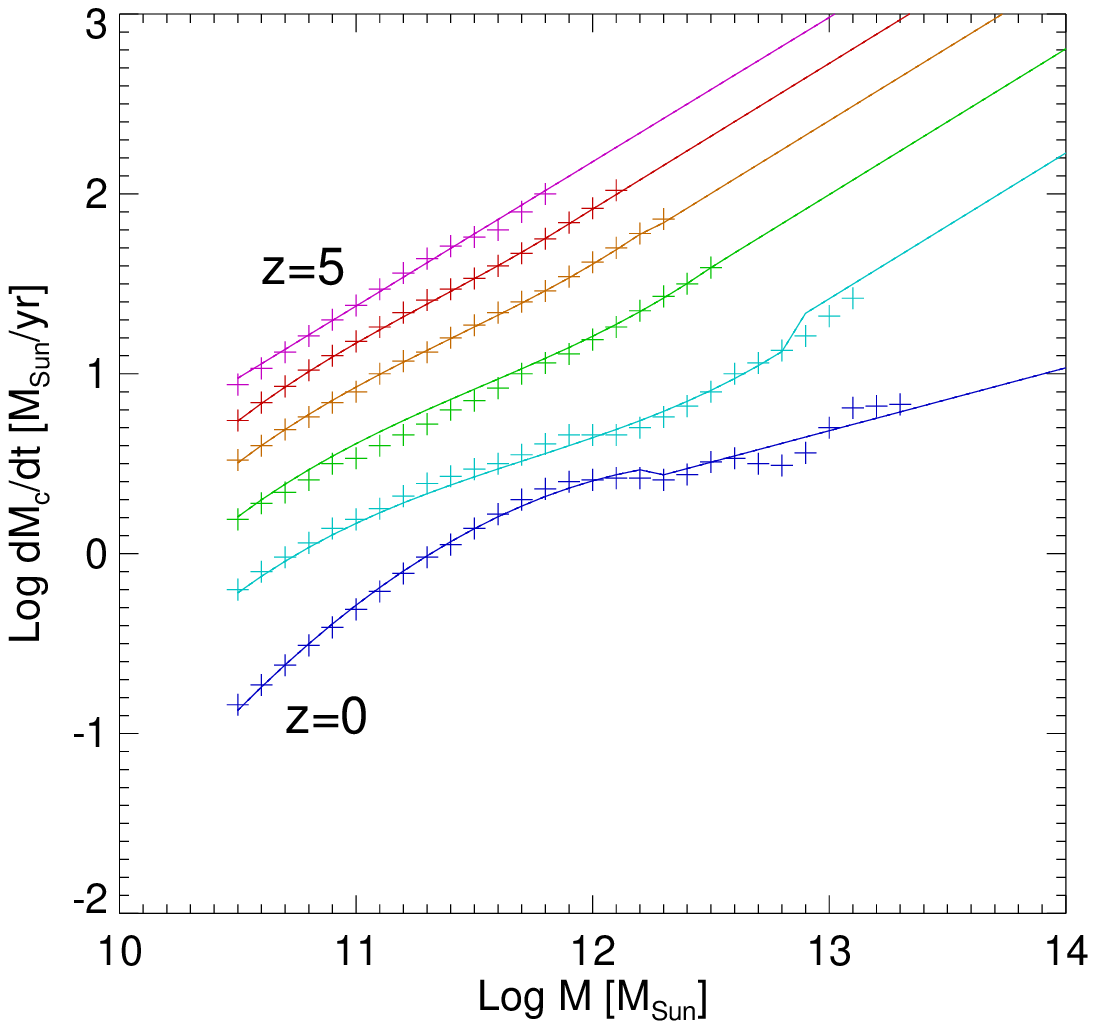}
\end{center}
\caption{Cold gas accretion rates at $R_{\rm vir}$, as calculated in \citet{Voort11} (full lines)
and our approximation (dashed lines). The results are shown at redshifts $z\sim 0,1,2,3,4,5$, from the lowest to the highest line (dark blue, light blue, green, orange, red, rose, respectively).
}\label{fig:A1}
\end{figure}

\begin{table*}
\begin{minipage}{150mm}
\caption{Our interpolation and extrapolation of FG11 cold gas
accretion rates at 1, 0.5 and 0.2 $R_{\rm vir}$ and at $z\geq 1$. Given are the radius, redshift
range, mass range, and the used shape of a polynomial (as a
function of $z'=\log(1+z)$ and $M'=\log M$). Here $M_{\rm
lim,1}=13.1-0.3z$ and $M_{\rm lim,2}=12.4-0.2z$. Polynomials
represent $\log\dot{M_{\rm c}}$.} \label{symbols}
\begin{tabular}{@{}lcccc}
\hline radius $[R_{\rm vir}]$ & $z$ range & $\log M$ range & polynomial $p(z',M')$ \\ %% & coefficients \\
\hline           1        & $1-5$     &  $>M_{\rm lim,1}$   &  $p(M',z')=a_{0}+a_{1}z'+0.81M'$ $^{*1}$  \\
                 1        & $1-5$     &  $<M_{\rm lim,1}$   &  $p(M',M'^{2},M'^{3},z'M',z'M'^{2},z'M'^{3})$  \\
%%                 1        & $<1$      & $>12.2$        &  $p(M')$   \\
%%                 1        & $0$      & $<12.2$        &  $p(M',M'^{2})$  \\
%%                 1        & $<1$     & $<12.2$        &  lin. int.$^{*2}$ between $z=0$ and $z=1$ \\
                 1        & $>5$      & $10-15$        &  $p(M',z',z'M')$  \\
\hline         0.5        & $1-5$     &  $>M_{\rm lim,2}$   &  $p(M',z')=a_{0}+a_{1}z'+0.81M'$  \\
               0.5        & $1-5$     &  $<M_{\rm lim,2}$   &  $p(M',M'^{2},M'^{3},z'M',z'M'^{2},z'M'^{3})$  \\
%%               0.5        & $0$      & $>12.2$        &  const   \\
%%               0.5        & $<1$      & $>12.2$        &  lin. int. between $z=0$ and $z=1$   \\
%%               0.5        & $0$      & $<12.2$        &  $p(M',M'^{2})$  \\
%%               0.5        & $<1$     & $<12.2$        &  lin. int. between $z=0$ and $z=1$ \\
               0.5        & $>5$      & $10-15$        &  $p(M',z',z'M')$  \\
\hline    %%     0.2        & $0$      & $<12.3$         &  $p(M',M'^{2})$  \\
          %%     0.2        & $0$      & $>12.3$         &  const  \\
               0.2        & $1,2,3,4$ & $<11$          &  $p(M')$  \\
               0.2        & $1,2,3,4$ & $>11$          &  $p(M')$  \\
               0.2        & $5$       & $<11.2$        &  $p(M')$  \\
               0.2        & $5$       & $>11.2$        &  $p(M')$  \\
               0.2        & $1-5$     & $10-15$        &  lin. int. between 2 nearest integer redshifts \\
               0.2        & $>5$      & $10-15$        &  lin. extrapolation from $z=4$ and $z=5$ \\
\hline
\end{tabular}
\medskip
\label{tab:4}

$^{*1}$ - coefficient 0.81 is used from \citet{FG11} fit
for
cold gas + interstellar medium accretion rates
at $z\geq2$.
For $M>M_{\rm lim}$ at redshifts $z=1-5$ and at radii $R_{\rm vir}$ and
$0.5R_{\rm vir}$ we use fit that has the same slope as that one of
FG11. For each redshift ($z=1,2,3,4,5$) we find the point from
the plot in the middle between the two points that are related to
the most massive haloes: $\log M=\frac{1}{2}(\log M (1)+\log M
(2))$, $\log \dot{M}_{c}=\frac{1}{2}(\log \dot{M}_{\rm c} (1)+\log
\dot{M}_{\rm c} (2))$, and fit these 6 points with straight lines
with shape $p(M',z')=a_{0}+a_{1}z'+0.81M'$.

$^{*2}$ - linear interpolation
\end{minipage}
\end{table*}

\subsection{Cold gas accretion rates in \citet{Voort11}}\label{sec:X.2}

We approximate cold gas accretion rates calculated in \citet{Voort11}, $\dot{M}_{\rm c,vdv}$,
as a function of cold gas accretion rates calculated in \citet{FG11}, $\dot{M}_{\rm c,fg}$, as follows:
for masses $M>M_{\rm lim,3}$: $\log\dot{M}_{\rm c,vdv} = \log\dot{M}_{\rm c,fg} + 0.33$,
and for masses $M<M_{\rm lim,3}$: $\log\dot{M}_{\rm c,vdv} = \log\dot{M}_{\rm c,fg} + (-4.83 +0.42\log M +0.41 \log(1+z))$,
where $\log M_{\rm lim,3}=12.5-0.3 z$ for $0.5<z<4$,
$\log M_{\rm lim,3}= 12.2$ for $z<0.5$, and
$\log M_{\rm lim,3}= 11.5$ for $z>4$.

\begin{figure}
\begin{center}
  \includegraphics*[width=8cm]{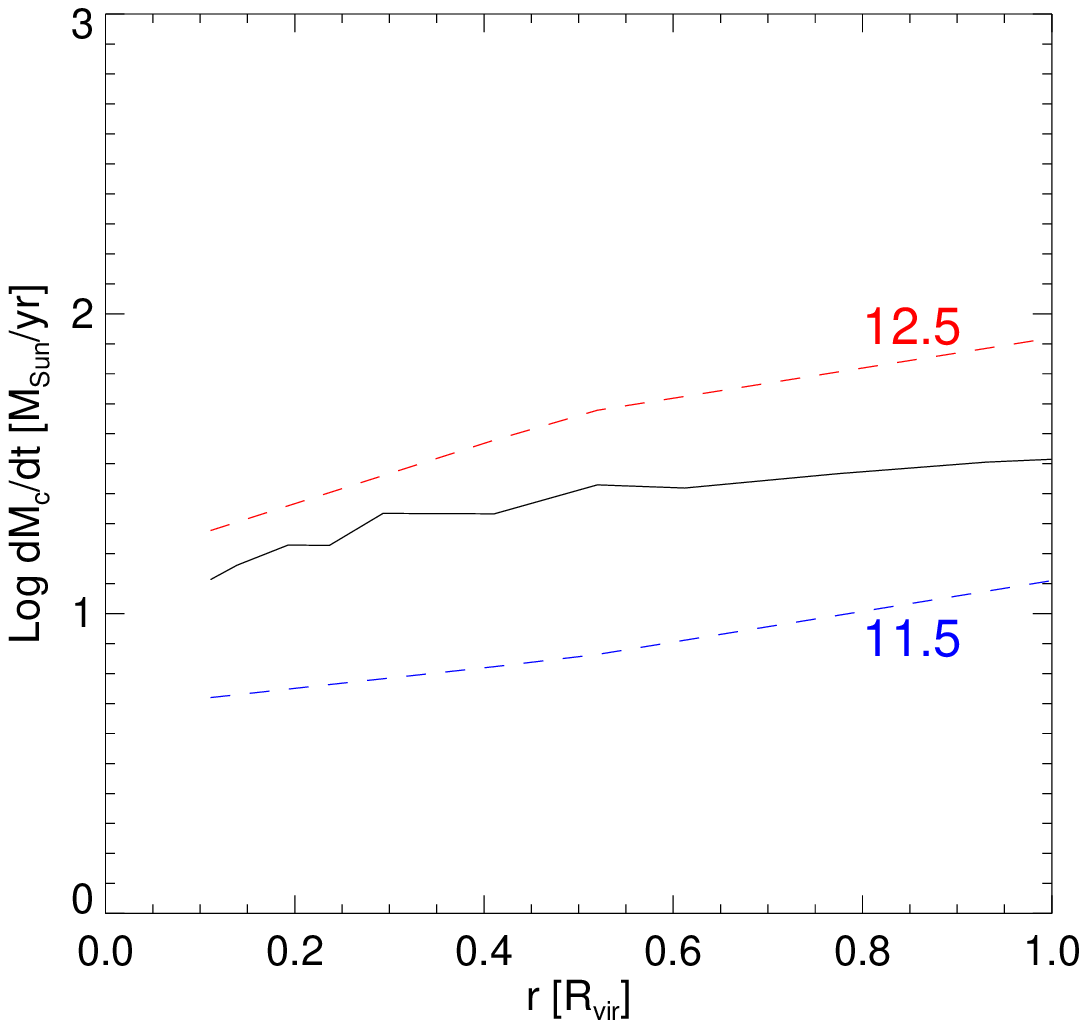}
\end{center}
\caption{Cold gas accretion rates as a function of radius inside haloes at $z=2$ and $10^{11.5} M_{\odot} < M_{\rm halo} < 10^{12.5} M_{\odot}$ as calculated in \citet{Voort12} (full line), their fig. 3. Dashed blue and red line denote our approximation at masses $10^{11.5} M_{\odot}$ and $10^{12.5} M_{\odot}$, respectively.
}\label{fig:A2}
\end{figure}

Figure \ref{fig:A1} shows
that in this way we could well describe $\dot{M}_{\rm c,vdv}$.
We assume that the same relation between $\dot{M}_{\rm c,vdv}$ and $\dot{M}_{\rm c,fg}$ holds at each radius $r$ inside a halo.
In Figure \ref{fig:A2} we show that
at $z=2$ and for halo masses $10^{11.5} M_{\odot} < M_{\rm halo} < 10^{12.5} M_{\odot}$
our approximation of $\dot{M}_{\rm c,vdv}$
is in agreement with \citet{Voort12} cold gas accretion rates
at radii $r\sim 0.1-1 R_{\rm vir}$ inside a halo.

%%\subsection{Resolution}\label{sec:X.3}
\subsection{Resolution of the DM simulation}\label{sec:6.4}

As Figure \ref{fig:8} shows the discrepancy at $SFR=10 M_{\odot}$ yr$^{-1}$,
we compare our results from the DM simulation with the results calculated from the Millennium-II simulation \citep[][]{Mill1,Mill2}, which has more than 100 times better mass resolution and $2-3$ times smaller volume.
  Millennium-II simulation is a pure dark matter simulation, which is run in a periodic box of size
$100$ Mpc $h^{-1}$, using cosmological parameters $(\Omega_{m},\Omega_{\Lambda},h)=(0.25,0.75,0.73)$.
It uses $10^{10}$ particles with masses $6.9\times 10^{6} h^{-1} M_{\odot}$.

\begin{figure}
\begin{center}
  \includegraphics*[width=8cm]{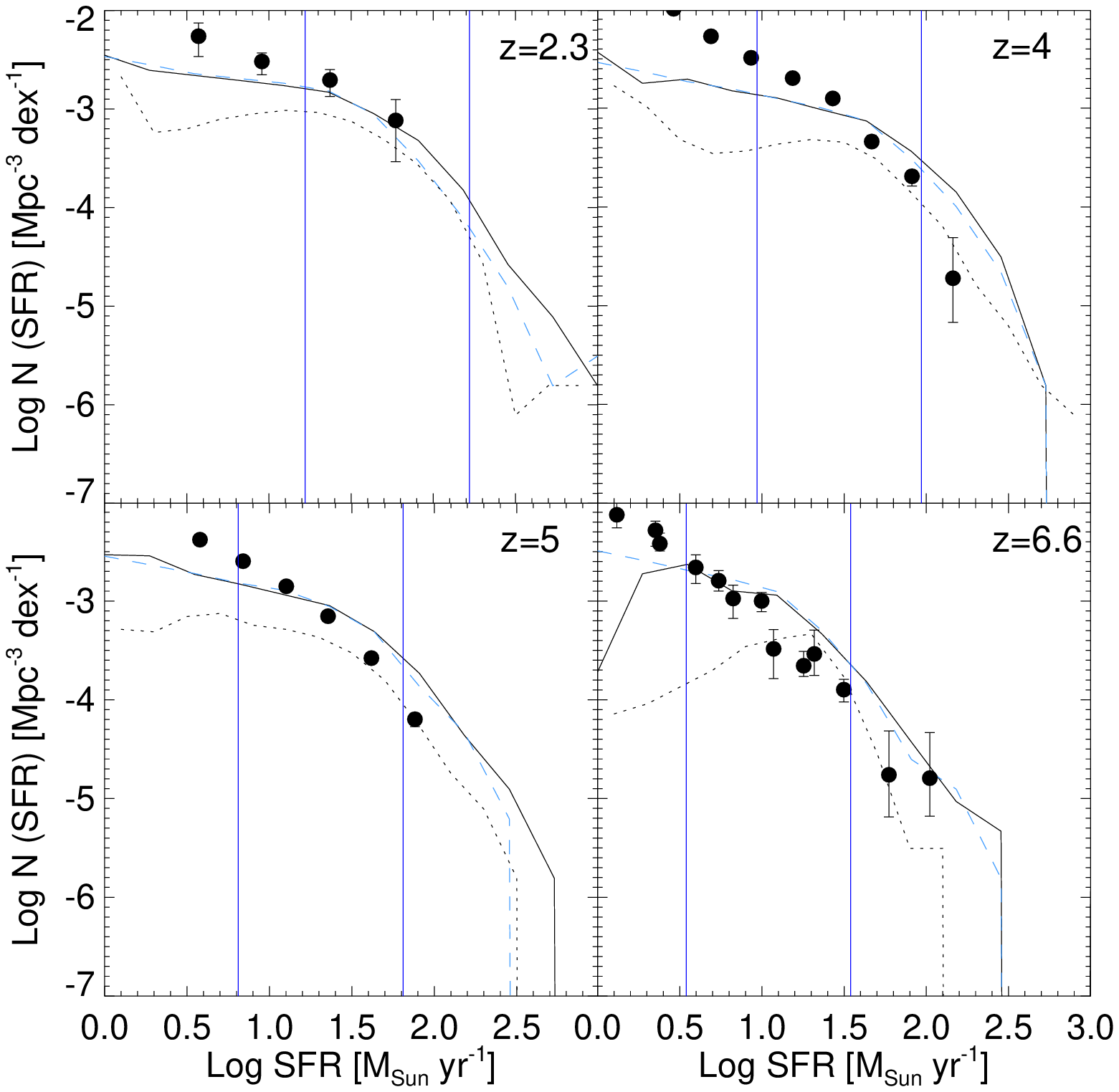}   %%Fig18
\end{center}
\caption{Star formation rate functions from our model at redshifts $z\sim2.3,4,5,6.6$
(dotted lines)
and from Millennium-II simulation.
SFRFs are calculated from the Millennium-II simulation for
total subhalo masses (light blue dashed lines) and for subhalo masses m-crit200 (black full lines; see the text).
Other notation is the same as in the Figure \ref{fig:8}.
}\label{fig:18}
\end{figure}

\begin{figure}
\begin{center}
  \includegraphics*[width=8cm]{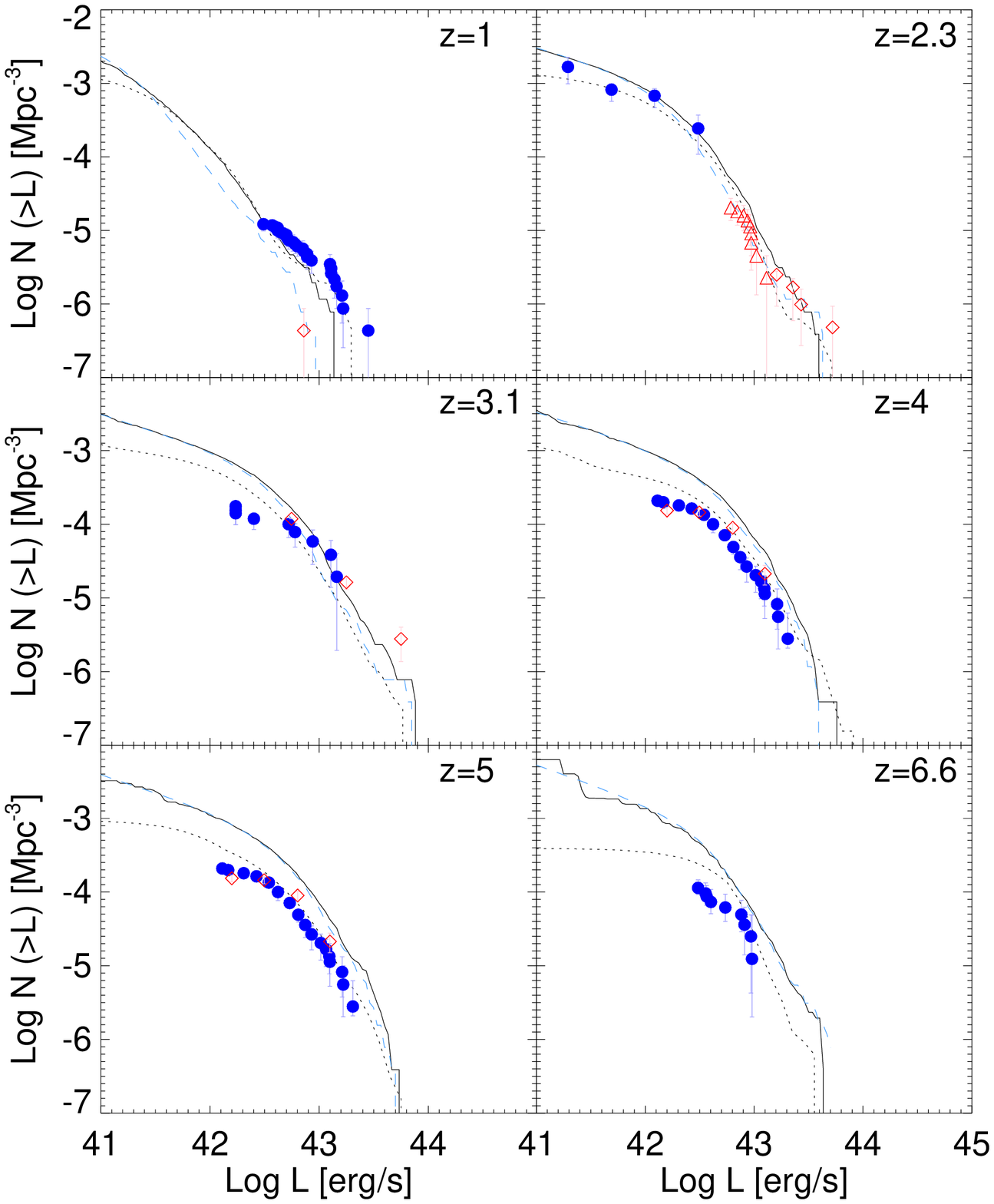}  %%Fig19
\end{center}
\caption{%%Cumulative luminosity functions (LFs)
Cumulative luminosity functions at a range of redshifts.
Notations are the same as in the Figure \ref{fig:4},
except that now full and dashed lines represent LFs
calculated from the Millennium-II simulation, and dotted lines are LFs from DM simulation.
Full and dashed lines represent
LFs calculated for subhalo masses m-crit200 and for total subhalo masses, respectively.
}\label{fig:19}
\end{figure}

\begin{figure}
\begin{center}
  \includegraphics*[width=8cm]{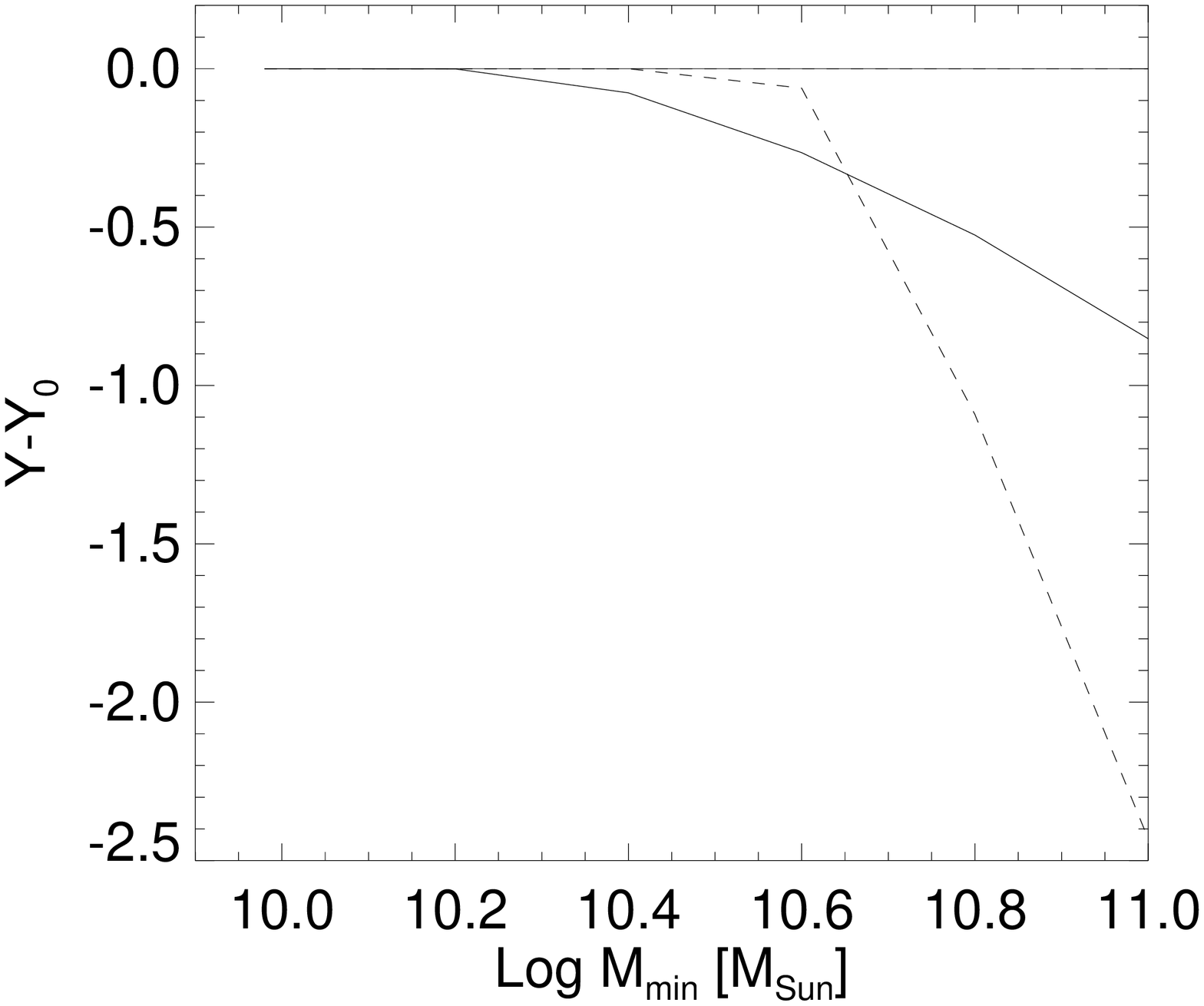}   %%Fig22
\end{center}
\caption{Convergence properties of LFs and SFRFs.
Maximum and minimum differences of LFs (full lines) and SFRFs (dashed lines) %% and SFRDs (dotted lines)
 between the values calculated for minimum halo masses of
  $M_{\rm min}$ and $\sim 10^{10} M_{\odot}$, as a function of $M_{\rm min}$.
  Minimum differences are equal to $\sim 0$ for both LFs and SFRFs.
}\label{fig:22}
\end{figure}

At each redshift, we select subhaloes with more than 1000 particles (which corresponds to a mass of $9.45\times 10^{9} M_{\odot}$).
For each subhalo we select all of its progenitors from the previous snapshot with more than 50 particles.
For all subhaloes we select its total mass. For subhaloes which are dominant in its friend-of-friends group, we select its mass m-crit200, which is the mass within the radius where the subhalo has an overdensity 200 times the critical density of the simulation. 

Using these data, we calculate LFs and SFRFs %%and mass functions, 
in the same way as for DM simulation (Figures \ref{fig:18} and \ref{fig:19}, respectively).
When we use the Millennium-II simulation, the results are almost the same, except that at low end SFRFs are higher and in a relatively good agreement with observations.
LFs almost did not change at the range of luminosities in which LABs and LAEs from Figure \ref{fig:4} are observed.

%%***
%%From the Figures  \ref{fig:20}, one can see that mass functions are similar for both simulations, except at low end,
%%where the Millennium-II simulation, which has better mass resolution, shows higher values, and at high redshift,
%%where in the Millennium-II simulation there are more massive (and luminous) haloes.
%%Similar holds for the figures \ref{fig:18} and \ref{fig:19}.

Figure \ref{fig:22} shows the convergence properties of LFs (for luminosities in the range $[10^{41},10^{44}]$ erg s$^{-1}$)
and SFRFs (for SFRs in the range $[1,1000]$ $M_{\odot}$ yr$^{-1}$). %% and SFRDs.
It could be seen that for mass resolutions $M_{\rm min} \sim 10^{10.4} M_{\odot}$ and
$M_{\rm min} \sim 10^{10} M_{\odot}$ the calculated LFs and SFRFs
almost do not differ.
Next, we examine convergence properties as following.
We define an array with minimum halo masses $r = \log M_{\rm min}=[10.6,10.4,10.2,10]$, and an array of
differences between a value of a LF(L,z) calculated for $r_{i}$ and $r_{i+1}$:
$b_{i}=LF(L,z,r_{i+1})-LF(L,z,r_{i})$.
For the array $b$ it holds $0<b_{i}<0.15^{i}$, for $i =$ 0, 1, 2.
If the same inequality holds for $i \geq 3$, then $LF(L,z,r_{i})<LF(L,z,r_{2})+0.15^{3}+...+0.15^i$.
This implies that $LF(L,z,r_{i})$ and $LF(L,z,r_{2})$ would not differ by more than
$0.15^{3}+...+0.15^i=(0.15^{3}-0.15^{i+1})/0.85$, which is less than 0.01.
The same holds for SFRF(SFR,z).

\end{document}